\documentclass[
a4paper,
aps,
pra,
showpacs,
twocolumn,
nosuperscriptaddress]
{revtex4-2} 

\usepackage[utf8]{inputenc}  
\usepackage[T1]{fontenc}     
\usepackage[american]{babel}  
\usepackage[svgnames,dvipsnames]{xcolor}
\usepackage{CJKutf8}
\usepackage{utfsym}
\usepackage[colorlinks,
citecolor=DarkBlue,
linkcolor=DarkBlue,
linktocpage,
unicode]{hyperref}
\usepackage{amsmath,amssymb,amsthm,bm,mathtools,amsfonts,mathrsfs,bbm,dsfont} 

\newtheorem{theorem}{Theorem}

\newcommand\norm[1]{\left\lVert#1\right\rVert}

\usepackage{braket}
\usepackage[normalem]{ulem}
\usepackage{comment}
\newcommand{\ketbra}[2]{\ensuremath{| #1 \rangle\!\langle #2 |}}
\usepackage{url}

\begin{document}
\title{Communication advantages of a quantum switch of two pure-loss channels}

\author{Carlo Marconi}
\affiliation{Istituto Nazionale di Ottica - Consiglio Nazionale delle Ricerche (INO-CNR), Largo Enrico Fermi 6, 50125 Firenze, Italy}

\date{\today}
\begin{abstract}
\noindent The quantum switch can enable communication through otherwise unusable channels, a phenomenon known as causal activation. We show that a switch of two identical pure-loss channels can achieve strictly greater energy-constrained classical and quantum capacities than their sequential use, even when two-way classical communication is available. Remarkably, the switch also activates quantum communication over a range of transmissivities for which the sequential channel has zero quantum capacity.
\end{abstract}

\maketitle


\textit{Introduction}.--- Photon loss represents a fundamental limitation in quantum communication, severely restricting the transmission of quantum information through optical channels. This process is mathematically described by a bosonic pure-loss channel \cite{weedbrook2012gaussian}, whose unassisted quantum capacity — the maximum rate at which quantum information can be reliably transmitted — vanishes when its transmissivity is below a certain threshold \cite{wolf2007quantum}. This limitation motivates the development of strategies to enable quantum communication in regimes where direct transmission is not possible.

In recent years, growing interest has focused on processes in which quantum channels are combined in a superposition of alternative configurations \cite{kristjansson2020resource}. A paradigmatic example is the quantum switch \cite{chiribella2013quantum}, which coherently controls the order in which two channels act on an input state. The quantum switch has been shown to enable the transmission of classical information through completely depolarizing channels \cite{ebler2018enhanced} and even to activate the quantum capacity of entanglement-breaking channels \cite{chiribella2021indefinite}. Whether these advantages extend to bosonic pure-loss channels remains an important open question, particularly because two pure-loss channels commute under sequential composition, suggesting that a coherent superposition of their orders might offer no advantage.

Here, we show that the quantum switch can nevertheless outperform sequential transmission under an input energy constraint. We derive an exact representation of the resulting channel and compare its ability to transmit classical and quantum information with that of two identical pure-loss channels applied sequentially. We prove that the quantum switch has a strictly greater classical capacity than the corresponding sequential channel. We further establish an advantage in quantum capacity, including when two-way classical assistance is available. Remarkably, in the unassisted setting, we identify a regime in which the quantum switch enables quantum information transmission even though the corresponding sequential channel has zero quantum capacity. 
Our results illustrate the potential of indefinite causal order for enhancing quantum communication through lossy optical channels.

\textit{Quantum switch}.--- Given two quantum channels $\mathcal{E}$ and $\mathcal{F}$, the action of a quantum switch over a composite system $\rho_A \otimes \omega_C$ reads \cite{chiribella2013quantum,ebler2018enhanced}
\begin{equation}
\label{eq:switch_formal}
    \mathcal{S}(\mathcal{E}, \mathcal{F})(\rho_A \otimes \omega_C) := \sum_{i,j} S_{ij} (\rho_A \otimes \omega_C) S_{ij}^{\dagger}~.
\end{equation}
Here the Kraus operators $S_{ij}$ are given by
\begin{equation}
    \label{eq:switch_kraus_ops}
    S_{ij} := E_{i} F_{j} \otimes \ketbra{0}{0} +  F_{j} E_{i} \otimes \ketbra{1}{1}~,
\end{equation}
where $E_{i}$ and $F_{j}$ denote the Kraus operators of the quantum channels $\mathcal{E}$ and $\mathcal{F}$, respectively.

In what follows, we restrict to pure-loss channels $\mathcal{E}_{\lambda}$, which are used to model the transmission of photons across a medium. Formally, their action can be described in terms of a beamsplitter with transmissivity $\lambda \in [0,1]$, which mixes an input state $\rho_{A}$ with the vacuum state of the environment, i.e., \cite{mele2024quantum}
\begin{equation}
    \mathcal{E}_{\lambda}(\rho_A) = \mbox{Tr}_{E}\big[U_{\lambda}(\rho_{A} \otimes \ketbra{0}{0}_{E})U_{\lambda}^{\dagger}\big]~,
\end{equation}
with $U_{\lambda} = \exp[\arccos{(\sqrt{\lambda})}(\hat{a}^{\dagger} \hat{e} - \hat{a} \hat{e}^{\dagger})]$, where $\hat{a}$ and $\hat{e}$ denote the bosonic annihilation operators of the system $A$ and the environment $E$, respectively. 

\noindent Note that a pure-loss channel $\mathcal{E}_\lambda$
admits Kraus operators of the form
\cite{mele2024quantum}
\begin{equation}
     K_{m}^{\lambda} = \sqrt{\frac{(1-\lambda)^m}{m!}}\lambda^{\hat{n}/2} \hat{a}^{m}~.
\end{equation}

We assume that the control qubit $C$ is initialized in a pure state and is inaccessible to the sender. For a fixed control state $\omega_C$, the quantum switch therefore induces an effective quantum channel, denoted $\mathcal{S}(\rho_A) := \mathcal{S}(\mathcal{E},\mathcal{F})(\rho_A \otimes \omega_C)$,  from Alice to Bob, who receives both the bosonic output $B$ and the control qubit $C$ (see, e.g., \cite{ebler2018enhanced}). In what follows, we use the term ``quantum switch'' to refer to this latter channel. Throughout, we choose $\omega_C = \ketbra{+}{+}$, with
$\ket{+}=(\ket{0}+\ket{1})/\sqrt{2}$, and restrict to two pure-loss channels with equal transmissivity. For ease of notation, we omit the control state and denote the resulting quantum switch by $\mathcal{S}_{\lambda}(\rho_A):= \mathcal{S}(\mathcal{E}_{\lambda},\mathcal{E}_{\lambda})(\rho_A \otimes \ketbra{+}{+})$.
When $\rho_A = \ketbra{\alpha}{\alpha}$, the action of the quantum switch reads
\begin{equation}
\label{eq:res1}
   \mathcal{S}_{\lambda}(\ketbra{\alpha}{\alpha}) = \ketbra{\lambda \alpha}{\lambda \alpha} \otimes 
    \frac{1}{2}
    \begin{bmatrix}
    1  &  e^{-\kappa_{\lambda} |\alpha|^2}\\
     e^{- \kappa_{\lambda} |\alpha|^2}  & 1
    \end{bmatrix}~,
\end{equation}
where $\kappa_\lambda := (1-\lambda)(1-\sqrt\lambda)^2 $.

\noindent The derivation of Eq.~(\ref{eq:res1}) is given in Appendix \ref{ap:res1}, along with the case of a generic input state. Note that tracing out the control qubit $C$ recovers the sequential composition of the two pure-loss channels, i.e., $\mbox{Tr}_C \circ \mathcal{S}_{\lambda}= \mathcal{E}_{\lambda^2} = \mathcal{E}_{\lambda} \circ \mathcal{E}_{\lambda}$. Any protocol for the sequential channel can therefore be implemented using the switch by discarding $C$ after each use.
Since this reproduces the sequential channel without changing Alice's inputs or their energy cost, every classical or quantum rate achievable with the sequential channel is also achievable with the switch under the same input energy constraint
\cite{khatri2020information}. Interestingly, we show that retaining $C$ yields a strict enhancement of each capacity over sequential use in a respective ranges of transmissivities.

\textit{Enhancement of the classical capacity}.--- 
The classical capacity of a quantum channel is the supremum of rates, in bits per channel use, at which classical information can be transmitted with vanishing decoding error. For some channels, including pure-loss channels with non-zero transmissivity, this capacity is infinite. We therefore impose a constraint on the mean input photon number and consider the energy-constrained classical capacity.

For a channel $\mathcal{L}:A\to B$, we first introduce the energy-constrained Holevo information $\chi_N(\mathcal{L})$, given by
\begin{equation}
\label{eq:def_holevo}
    \chi_{N}(\mathcal{L}):= \sup_{\{p_x, \rho_A^{x}\}} I(X;B)_{\sigma}~,
\end{equation}
where the subscript $N$ indicates that the optimization is restricted to discrete input ensembles such that the average state $\bar{\rho}_A:= \sum_{x} p_x\rho_A^{x}$ satisfies $\mbox{Tr}[\hat{n}_A\bar{\rho}_{A}]\leq N$.
When the average output entropy is finite, the mutual
information $I(X;B)_{\sigma}$ is defined as 
\begin{equation}
\label{eq:def_mutual}
     I(X;B)_{\sigma}:= H(B)_{\mathcal{L}(\bar{\rho}_A)}  - \sum_{x} p_x H(B)_{\mathcal{L}(\rho_A^{x})}~,
\end{equation}
where $\sigma := \sum_{x\in \mathcal{X}}p_x \ketbra{x}{x} \otimes \mathcal{L}(\rho_A^{x})$, $\{\ket{x}_X\}_{x\in\mathcal X}$ is an orthonormal
basis of the classical register $X$ and $H(X)_{\rho}=-\mbox{Tr}(\rho \log_2 \rho)$ denotes the von Neumann entropy of a system $X$ in a state $\rho$.
Thus, the energy-constrained classical capacity is given by \cite{holevo2003entanglement}
\begin{equation}
    \mathcal{C}_N(\mathcal{L}) = \lim_{n \to \infty} \frac{1}{n} \chi_{nN}(\mathcal{L}^{\otimes n})~,
\end{equation}
where the optimization is performed over joint input states $\rho_{A^{n}}$ whose average satisfies $\operatorname{Tr}[(\sum_{j=1}^{n} \hat{n}_{A_j}) \bar{\rho}_{A^n}] \leq nN$.

For a pure-loss channel with $\lambda>0$, it is known that the
energy-constrained classical capacity is achieved by a
Gaussian ensemble of coherent states whose average corresponds to a
thermal state \cite{giovannetti2004classical}. Its expression reads
\begin{equation}
    \mathcal{C}_{N}(\mathcal{E}_{\lambda}) = g(N \lambda)~,
\end{equation}
with $g(x) := (x+1) \log_2 (x+1) - x \log_2 (x)$. 

To establish a strict enhancement over the sequential composition of two pure-loss channels of transmissivity $\lambda$, we use the same Gaussian ensemble as input to the switch. For $N>0$, we draw the amplitudes of the coherent states $\rho_A^\alpha=\ketbra{\alpha}{\alpha}$ from the Gaussian distribution $p_N(\alpha)=e^{-|\alpha|^2/N}/(\pi N)$. Note that in this case, since $\alpha \in \mathbb{C}$ varies over a continuous alphabet $\mathcal{X}$, we need to replace the mutual information of Eq.~(\ref{eq:def_mutual}) with the Holevo quantity of a generalized ensemble \cite{holevo2006continuous}. This corresponds to the choice of a Borel probability measure $\pi$ on $\mathcal{X}$, together with a measurable family of states $\{\rho_A^x\}_{x \in \mathcal{X}}$ \cite{shirokov2025upper}. 
Hence, the Holevo quantity $\chi_{N}(\mathcal{L}(\pi))$ is given by \cite{holevo2006continuous}
\begin{equation}
    \label{eq:holevo_quantity}
    \chi_{N}(\mathcal{L}(\pi)):= \int_{\mathcal{X}} D \big(\mathcal{L}(\rho_A^x) || \mathcal{L}(\bar{\rho}_A(\pi)) \big) \pi(dx)~,
\end{equation}
where $\bar{\rho}_A(\pi):= \int_{\mathcal{X}} \rho_A^x \pi(dx)$ and $D(\rho||\sigma)$ denotes the quantum relative entropy. For a discrete ensemble, this expression reduces to Eq.~(\ref{eq:def_mutual}). Optimizing over all generalized ensembles satisfying the same mean-energy constraint yields $\chi_N(\mathcal L)$
\cite{holevo2006continuous}.

In our case, a generalized ensemble is given by $\pi(d\alpha):=p_N(\alpha)d^2\alpha$, together with the states $\{\rho_A^\alpha:= \ketbra{\alpha}{\alpha}\}_{\alpha\in\mathbb{C}}$.  We evaluate Eq.~(\ref{eq:holevo_quantity}) for the switch and obtain
\begin{equation}
\label{eq:res2}
 \chi_{N}(\mathcal{S}_{\lambda}(\pi))
=\mathcal{C}_N(\mathcal{E}_{\lambda^2})+\Delta_N~,
\end{equation}
where the explicit expression for $\Delta_N$ is given in Appendix~\ref{ap:res2}, along with the derivation of Eq.~(\ref{eq:res2}). Since the chosen ensemble satisfies the input energy constraint,
$\chi_{N}(\mathcal{S}_{\lambda}(\pi))$ provides a lower bound on
$\mathcal{C}_N(\mathcal{S}_\lambda)$
\cite{holevo2003entanglement, holevo2006continuous}.
Therefore, $\Delta_N>0$ establishes a strict advantage over
the sequential channel. In Appendix~\ref{ap:res3}, we prove
that $\Delta_N>0$ for every $N>0$ and $0\leq\lambda<1$,
which implies
\begin{equation}
\label{eq:res3}
    \mathcal{C}_N(\mathcal{S}_\lambda)
    >\mathcal{C}_N(\mathcal{E}_{\lambda^2})~,
   \quad 0\leq\lambda<1~.
\end{equation}

\textit{Enhancement of the quantum capacity}.--- We now turn to the quantum capacity and prove that, also in this case, the switch allows for an enhancement when compared to the sequential use of two pure-loss channels. 

\noindent We first introduce the energy-constrained coherent information of a channel $\mathcal{L}:A \to B$, i.e., \cite{sharma2018bounding}
\begin{equation}
    \label{eq:coherent_info}
    \mathcal{Q}^{(1)}_{N}(\mathcal{L}):= \sup_{\rho_A : \operatorname{Tr}(\rho_A \hat{n}_A) \leq N} I_c(\rho_A, \mathcal{L})~,
\end{equation}
with
\begin{equation}
    I_c(\rho_A, \mathcal{L}):= H(B)_{\mathcal{L}(\rho_A)}-H(E)_{\hat{\mathcal{L}}(\rho_A)}~.
\end{equation}
Here, $\hat{\mathcal{L}}:A \to E$ denotes a complementary channel of $\mathcal{L}$ and we assume all entropies to be finite.

\noindent Thus, the energy-constrained quantum capacity is defined as \cite{sharma2018bounding}
\begin{equation}
        \mathcal{Q}_{N}(\mathcal{L})= \lim_{n \rightarrow \infty} \frac{1}{n}  \mathcal{Q}^{(1)}_{nN}(\mathcal{L}^{\otimes n})~,
\end{equation}
where the optimization is taken over joint input states $\rho_{A^{n}}$ satisfying $\operatorname{Tr}[(\sum_{j=1}^{n} \hat{n}_{A_j})\rho_{A^n}] \leq nN$. 

The energy-constrained quantum capacity for a pure-loss channel $\mathcal{E}_{\lambda}$ reads \cite{wilde2012quantum} 
\begin{equation}
\label{eq:qc_pureloss}
\mathcal{Q}_N(\mathcal{E}_{\lambda}) = \max\{g(\lambda N) - g((1-\lambda) N),0\}~.
\end{equation}
For $\lambda\geq1/2$, this capacity is attained by a thermal input state $\tau_N$, while for $\lambda<1/2$, the channel is anti-degradable and its quantum capacity vanishes.

We show that the coherent information for the switch $\mathcal{S}_\lambda$ at input state $\tau_N$ is given by
\begin{equation}
\label{eq:res_coherent}
    I_c(\tau_N, \mathcal{S}_{\lambda}) = I_c(\tau_N, \mathcal{E}_{\lambda^2})+ \Delta_N^Q~,
\end{equation}
where the expression of $\Delta_N^{Q}$ is reported in Appendix~\ref{ap:res4}, along with the derivation of Eq.~(\ref{eq:res_coherent}). We further show that $\Delta_N^Q > 0$ for every $N>0$ and $0<\lambda<1$. Since the thermal input attains $\mathcal Q_N(\mathcal E_{\lambda^2})$ for $1/\sqrt{2} \leq \lambda < 1$, this implies
\begin{equation}
\label{eq:res4}
\mathcal{Q}_N(\mathcal{S}_\lambda)> \mathcal{Q}_N(\mathcal{E}_{\lambda^2})~, \quad 1/\sqrt2 \leq\lambda<1~.
\end{equation}

\textit{Activation of the quantum capacity}.---We observe that the thermal input state establishes a strict enhancement for $1/\sqrt{2}\leq\lambda<1$, corresponding to the region where $\mathcal{Q}_N(\mathcal{E}_{\lambda^2})=I_c(\tau_N,\mathcal{E}_{\lambda^2})\geq0$. Below this threshold, the sequential capacity vanishes, while the sequential coherent information at $\tau_N$ is negative. A positive $\Delta_N^Q$ is therefore not sufficient to establish a strict enhancement in this regime. Nevertheless, we can show that the switch has non-zero quantum capacity over a range of transmissivities for which the sequential channel has zero quantum capacity, thus demonstrating activation through indefinite causal order.

To prove this result, we restrict the switch input to the subspace spanned by the vacuum and single-photon states. Note that any code for the resulting channel with a qubit input can also be used with the full switch under the same energy constraint. In Appendix~\ref{ap:res5}, we construct a degrading map for this restricted channel and identify an input state with strictly positive coherent information that satisfies the energy constraint. This establishes a positive quantum communication rate for the switch where the capacity of the sequential channel vanishes. More formally, we find
\begin{equation}
\label{eq:res5}	
\mathcal{Q}_N(\mathcal{S}_\lambda) > 0 = \mathcal{Q}_N(\mathcal{E}_{\lambda^2})~, \quad \lambda_* <\lambda \leq 1/\sqrt2~,
\end{equation}
with $\lambda_* \simeq 0.70577576$.

\textit{Enhancement of the two-way quantum capacity}.---We conclude by considering the energy-constrained two-way assisted quantum capacity, $\mathcal{Q}_{2,N}$. This is the supremum of rates, in qubits per channel use, achievable with vanishing error when the sender and receiver can exchange unlimited noiseless classical messages in both directions, subject to a mean input photon number of at most $N$ per use \cite{davis2018energy}. We show that, even with this assistance, the switch retains an advantage over sequential transmission.

The energy-constrained two-way quantum capacity of a pure-loss channel with transmissivity $0<\lambda\leq1$ has recently been determined in \cite{pirandola2026energy,falco2026energy} and is given by
\begin{equation}
\mathcal{Q}_{2,N}(\mathcal{E}_{\lambda})=g(N)-g((1-\lambda)N)~.
\end{equation}

In Appendix~\ref{ap:res6}, we construct an achievable protocol for the switch. Alice prepares independent two-mode squeezed vacuum states, retaining one mode and sending the other through the switch. Bob measures the control qubit in the ${\ket{+},\ket{-}}$ basis and communicates the outcomes to Alice. The parties then group the resulting conditional states and distill entanglement using classical communication from Bob to Alice. The distilled entanglement can subsequently be used for teleportation, yielding an achievable quantum communication rate without additional uses of the switch. Averaging the achievable distillation rates over the
control outcomes gives a rate strictly greater than
$g(N)-g((1-\lambda^2)N)$, yielding
\begin{equation}
\label{eq:res6}
\mathcal{Q}_{2,N}(\mathcal S_\lambda)>\mathcal{Q}_{2,N}(\mathcal{E}_{\lambda^2})~, \quad  0<\lambda<1~.
\end{equation}

\textit{Conclusions}.--- We have shown that a quantum switch of two identical bosonic pure-loss channels can outperform their sequential composition under the same input energy constraint, despite the commutativity of the constituent channels. For every $N>0$, we establish a strict enhancement of the energy-constrained classical capacity for $0 \leq \lambda < 1$ and of the two-way-assisted quantum capacity for $0 < \lambda < 1$. For unassisted quantum communication, thermal inputs establish a strict advantage for
$1/\sqrt{2} \leq \lambda < 1$. Restricting the input to the vacuum and single-photon subspace additionally proves activation for $\lambda_* < \lambda \leq 1/\sqrt{2}$, with
$\lambda_* \simeq 0.70577576$, where the sequential channel has zero quantum capacity. To the best of our knowledge, this provides the first example of causal activation for a
continuous-variable channel, thus solving an open question originally posed in \cite{koudia2022deep}.

Several questions remain open. Determining the exact energy-constrained capacities requires identifying optimal input ensembles and communication protocols, together with
matching converse bounds. In particular, it remains to determine whether the Gaussian ensemble of coherent states is optimal for classical communication and in which regimes thermal inputs are optimal for quantum communication. Another direction is to investigate whether more general switched configurations, including superpositions of additional channel orders, allow to extend the activation of the quantum capacity at lower transmissivities. Finally, extending the analysis to channels subject to both loss and dephasing would clarify the robustness of the capacity advantage and whether dephasing affects the gap relative to the corresponding noisy sequential channel.

\textit{Acknowledgments \& AI statement}.--- 
The author conceived the original idea for this work. OpenAI’s ChatGPT-6.1 Sol was used to assist with the derivation of the mathematical results and bibliography research. The author independently reviewed and checked all the results and takes full responsibility for the content of this work. The author acknowledges support from QUINTET - Quantum Internet Technologies, CUP B53C26001590005.

\onecolumngrid
\appendix

\section{Proof of Eq.~(\ref{eq:res1})} 
\label{ap:res1}
In this Appendix, we derive the action of a quantum switch of two pure-loss channels for a generic input state $\rho_{A}$. Let us recall that the Kraus operators of a pure-loss channel take the form \cite{mele2024quantum}
\begin{equation}
\label{eq:kraus_single}
    K_{i}^{\lambda} = \sqrt{\frac{(1-\lambda)^i}{i!}}\lambda^{\hat{n}/2} \hat{a}^{i}~,
\end{equation}
from which, using Eqs.~(\ref{eq:switch_formal})-(\ref{eq:switch_kraus_ops}) in the main text, we find
\begin{align}
    \label{eq:switch_kraus}
     \mathcal{S}_\lambda(\rho_A) &= \frac{1}{2}\sum_{i,j=0}^{\infty} K_{i}^{\lambda} K_{j}^{\lambda} \rho_A (K_{j}^{\lambda})^{\dagger} (K_{i}^{\lambda})^{\dagger} \otimes \ketbra{0}{0} + 
     \frac{1}{2}\sum_{i,j=0}^{\infty} K_{i}^{\lambda} K_{j}^{\lambda} \rho_A (K_{i}^{\lambda})^{\dagger} (K_{j}^{\lambda})^{\dagger} \otimes \ketbra{0}{1} \\ 
     \nonumber
     &+\frac{1}{2}\sum_{i,j=0}^{\infty} K_{j}^{\lambda} K_{i}^{\lambda} \rho_A (K_{j}^{\lambda})^{\dagger} (K_{i}^{\lambda})^{\dagger} \otimes \ketbra{1}{0} +
     \frac{1}{2}\sum_{i,j=0}^{\infty} K_{j}^{\lambda} K_{i}^{\lambda} \rho_A (K_{i}^{\lambda})^{\dagger} (K_{j}^{\lambda})^{\dagger} \otimes \ketbra{1}{1}~.
\end{align}

\noindent Exploiting the Kraus representation of each channel, the above expression can be cast in the computational basis of the control qubit as
\begin{align}
    \mathcal{S}_\lambda(\rho_A) &= \frac{1}{2}
    \begin{bmatrix}
    \label{eq:switch_mat_1}
     \left(\mathcal{E}_{\lambda} \circ \mathcal{E}_{\lambda} \right)(\rho_A) &  \mathcal{N}(\rho_A)\\
     \mathcal{N}(\rho_A)^{\dagger} &  \left(\mathcal{E}_{\lambda} \circ \mathcal{E}_{\lambda} \right)(\rho_A)
     \end{bmatrix}\\
     &=\frac{1}{2}  \begin{bmatrix}
    \label{eq:switch_mat_2}
     \mathcal{E}_{\lambda^2}(\rho_A) &  \mathcal{N}(\rho_A)\\
     \mathcal{N}(\rho_A)^{\dagger} &  \mathcal{E}_{\lambda^2}(\rho_A)
    \end{bmatrix}~,
\end{align}
where we have used the fact that $\mathcal{E}_\lambda \circ \mathcal{E}_\lambda = \mathcal{E}_{\lambda^2}$ \cite{mele2024quantum} and we have set
\begin{equation}
    \mathcal{N}(\rho_A) := \sum_{i,j=0}^{\infty} K_{i}^{\lambda} K_{j}^{\lambda} \rho_A (K_{i}^{\lambda})^{\dagger} (K_{j}^{\lambda})^{\dagger}~.
\end{equation}

In order to find an explicit representation of Eq.~(\ref{eq:switch_mat_2}), let us first rewrite the product of two Kraus operators in Eq.~(\ref{eq:kraus_single}) as
\begin{align}
\label{eq:kraus_product}
    K_{i}^{\lambda} K_{j}^{\lambda} &= \sqrt{\frac{(1-\lambda)^{i+j}}{i! ~j!}}\lambda^{\hat{n}/2} \hat{a}^{i}\lambda^{\hat{n}/2} \hat{a}^{j}\\
    &= \sqrt{\frac{(1-\lambda)^{i+j}}{i! ~j!}}\lambda^{\hat{n}/2} \lambda^{\hat{n}/2} \lambda^{i/2}\hat{a}^{i} \hat{a}^{j}\\
    &= \sqrt{\frac{(1-\lambda)^{i+j}\lambda^{i}}{i! ~j!}}\lambda^{\hat{n}}\hat{a}^{i+j}~,
\end{align}
where we have used the fact that $\hat{a}^{i} \lambda^{\hat{n}/2} = \lambda^{\hat{n}/2}\lambda^{i/2} \hat{a}^{i}$. Recall that the Kraus operators $K^{\lambda^2}_{m}$ of the quantum channel $\mathcal{E}_{\lambda^2}$ are given by
\begin{equation}
    \label{eq:kraus_double}
    K_{m}^{\lambda^2} = \sqrt{\frac{(1-\lambda^2)^m}{m!}}\lambda^{\hat{n}} \hat{a}^{m}~.
\end{equation}
Hence, Eq.~(\ref{eq:kraus_product}) can be rewritten as
\begin{equation}
    K_{i}^{\lambda} K_{j}^{\lambda} =\sqrt{\binom{i+j}{i}\frac{\lambda^i}{(1+\lambda)^{i+j}}}K_{i+j}^{\lambda^2}~.
\end{equation}
Similarly, taking the product $ K_{j}^{\lambda} K_{i}^{\lambda}$ we find
\begin{equation}
     K_{j}^{\lambda} K_{i}^{\lambda} = \sqrt{\binom{i+j}{i}\frac{\lambda^j}{(1+\lambda)^{i+j}}}K_{i+j}^{\lambda^2}~,
\end{equation}
from which, setting 
\begin{equation}
    A_{i,j} := \sqrt{\binom{i+j}{i}\frac{\lambda^i}{(1+\lambda)^{i+j}}}~, \quad  B_{i,j}:=\sqrt{\binom{i+j}{i}\frac{\lambda^j}{(1+\lambda)^{i+j}}}~,
\end{equation}
Eq.~(\ref{eq:switch_mat_2}) takes the form
\begin{align}
\label{eq:switch_mat_3}
    \mathcal{S}_\lambda(\rho_A) &= \frac{1}{2}
    \sum_{i,j=0}^{\infty} K_{i+j}^{\lambda^2}\rho_A \left(K_{i+j}^{\lambda^2} \right)^\dagger \otimes
    \begin{bmatrix}
     A_{i,j}^2  & A_{i,j} B_{i,j}\\
      A_{i,j} B_{i,j} &   B_{i,j}^2
     \end{bmatrix}~.
\end{align}
Setting $m=i+j$, we find
\begin{align}
    \mathcal{S}_\lambda(\rho_A) &= \frac{1}{2}
    \sum_{m=0}^{\infty} K_{m}^{\lambda^2}\rho_A \left(K_{m}^{\lambda^2} \right)^\dagger \otimes
    \begin{bmatrix}
    \label{eq:switch_mat_4}
     \sum_{i=0}^{m} A^{2}_{i,m-i}  & \sum_{i=0}^{m}A_{i,m-i}B_{i,m-i}\\
      \sum_{i=0}^{m}A_{i,m-i}B_{i,m-i} &  \sum_{i=0}^{m} B^{2}_{i,m-i}
     \end{bmatrix}~.
\end{align}
Direct calculation shows that
\begin{align}
    \sum_{i=0}^{m}A^{2}_{i,m-i} = \sum_{i=0}^{m} \binom{m}{i}\frac{\lambda^{i}}{(1+\lambda)^m} = \frac{1}{{(1+\lambda)^m}}\sum_{i=0}^{m} \binom{m}{i}\lambda^{i} = 1~,
\end{align}
where in the last equality we have used the binomial theorem.
An analogous computation yields 
\begin{equation}
\sum_{i=0}^{m}B^{2}_{i,m-i} =1~.
\end{equation}
As for the off-diagonal terms in Eq.~(\ref{eq:switch_mat_4}), we find
\begin{equation}
    \sum_{i=0}^{m}A_{i,m-i}B_{i,m-i} = \sum_{i=0}^{m} \binom{m}{i}\frac{\lambda^{m/2}}{(1+\lambda)^m} = \frac{\lambda^{m/2}}{(1+\lambda)^m}\sum_{i=0}^{m} \binom{m}{i} = \left(\frac{2 \sqrt{\lambda}}{1+\lambda}\right)^m~,
\end{equation}
where in the last step we have used the relation $\sum_{i=0}^{m} \binom{m}{i} = 2^m$ and we have set 
$\gamma:= \frac{2 \sqrt{\lambda}}{1+\lambda} \in [0,1]$. 

\noindent Introducing the qubit state
\begin{equation}
\label{eq:sigma}
    \sigma_{m} = \frac{1}{2}
    \begin{bmatrix}
      1  & \gamma^{m}\\
      \gamma^{m} & 1
     \end{bmatrix}~,
\end{equation}
Eq.~(\ref{eq:switch_mat_4}) can be finally cast as
\begin{align}
\label{eq:switch_mat_5}
     \mathcal{S}_\lambda(\rho_A) &= 
    \sum_{m=0}^{\infty} K_{m}^{\lambda^2}\rho_A \left(K_{m}^{\lambda^2} \right)^\dagger \otimes
    \sigma_{m}~.
\end{align}

We now compute the action of the quantum switch for an input coherent state, i.e., $\rho_A = \ketbra{\alpha}{\alpha}$. Let us recall that the action of the Kraus operator $K_{m}^{\lambda^2}$ on a coherent state reads
\begin{align}
    K_{m}^{\lambda^2} \ket{\alpha} &= \sqrt{\frac{(1-\lambda^2)^{m}}{m!}} \lambda^{\hat{n}} \hat{a}^m \ket{\alpha} \\
    &= \sqrt{\frac{(1-\lambda^2)^{m}}{m!}} \alpha^{m} \lambda^{\hat{n}} \left(e^{-\frac{|\alpha|^{2}}{2}} \sum_{n=0}^{\infty} \frac{\alpha^{n}}{\sqrt{n!}} \ket{n}\right)\\
    &= \sqrt{\frac{(1-\lambda^2)^{m}}{m!}} \alpha^{m} e^{-\frac{(1-\lambda^2)|\alpha|^{2}}{2}} \left(e^{-\frac{\lambda^2|\alpha|^{2}}{2}} \sum_{n=0}^{\infty} \frac{(\lambda \alpha)^{n}}{\sqrt{n!}} \ket{n}\right)\\
    &= \sqrt{\frac{(1-\lambda^2)^{m}}{m!}} \alpha^{m} e^{-\frac{(1-\lambda^2)|\alpha|^{2}}{2}}  \ket{\lambda \alpha}~,
\end{align}
where we have used the fact that $\hat{a}^{m}\ket{\alpha} = \alpha^{m}\ket{\alpha}$. 
Substituting in Eq.~(\ref{eq:switch_mat_5}) yields
\begin{align}
    \mathcal{S}_\lambda(\ketbra{\alpha}{\alpha}) = \sum_{m=0}^{\infty}  
e^{-(1-\lambda^2)|\alpha|^2}
\frac{[(1-\lambda^2)|\alpha|^2]^m}{m!}
     \ketbra{\lambda\alpha}{\lambda\alpha} \otimes
    \sigma_{m} = \ketbra{\lambda\alpha}{\lambda\alpha} \otimes
    \sigma_C^{\alpha}~,
\end{align}
where we have introduced
\begin{equation}
    \sigma_C^{\alpha} := \sum_{m=0}^{\infty}   e^{-\mu}
\frac{\mu^m}{m!} \sigma_{m} =  \frac{1}{2}\sum_{m=0}^{\infty}   e^{-\mu}
\frac{\mu^m}{m!} 
    \begin{bmatrix}
      1  & \gamma^{m}\\
      \gamma^{m} & 1
     \end{bmatrix}~, \qquad \mu:=(1-\lambda^2)|\alpha|^2~.
\end{equation}
Note that the diagonal entries of $\sigma_C^{\alpha}$ are equal to $1/2$ since
\begin{equation}
   \sum_{m=0}^{\infty}   e^{-\mu}
\frac{\mu^m}{m!} =1~,
\end{equation}
being the normalization constraint of a Poisson distribution with parameter $\mu \geq 0$. The off-diagonal entries are easily found to be given by
\begin{align}
    \frac{1}{2}\sum_{m=0}^{\infty}   e^{-\mu}
\frac{\mu^m}{m!} \gamma^{m} =   \frac{1}{2}e^{-\mu} \sum_{m=0}^{\infty}  
\frac{(\gamma \mu)^m}{m!} = \frac{1}{2}e^{-\mu(1-\gamma)} = \frac{1}{2}e^{-(1-\lambda)(1-\sqrt\lambda)^2|\alpha|^2}~.
\end{align}

\noindent Finally, setting $\kappa_\lambda := (1-\lambda)(1-\sqrt\lambda)^2 $, the action of the quantum switch reads
\begin{equation}
\label{eq:action_switch}
   \mathcal{S}_{\lambda}(\ketbra{\alpha}{\alpha}) = \ketbra{\lambda \alpha}{\lambda \alpha} \otimes \sigma_C^{\alpha}~, \quad  
    \sigma_C^{\alpha}= \frac{1}{2}
    \begin{bmatrix}
    1  &  e^{-\kappa_{\lambda} |\alpha|^2}\\
     e^{- \kappa_{\lambda} |\alpha|^2}  & 1
    \end{bmatrix}~.
\end{equation}
This proves Eq.~(\ref{eq:res1}) in the main text.

\section{Proof of Eq.~(\ref{eq:res2})}
\label{ap:res2}
We compute the Holevo quantity $\chi_{N}(\mathcal{S}_{\lambda}(\pi))$ of Eq.~(\ref{eq:res2}) in the main text. Let us recall that the Gaussian probability distribution $p_N(\alpha)=\frac{e^{-|\alpha|^2/N}}{\pi N}$, together with the family of coherent states $\{\ketbra{\alpha}{\alpha}\}_{\alpha \in \mathbb{C}}$, defines a generalized ensemble $\{\pi:=p_N(\alpha) d^2\alpha, \ketbra{\alpha}{\alpha}\}_{\alpha \in \mathbb{C}}$  over $\mathbb{C}$, with $0<N<\infty$. 
Hence, Eq.~(\ref{eq:holevo_quantity}) in the main text reads
\begin{equation}
\label{ap:gen_ens}
    \chi_N\big(\mathcal{S}_{\lambda}(\pi)\big):=  \int d^2\alpha ~p_N(\alpha) D\bigl(\mathcal{S}_{\lambda}(\rho_A^{\alpha})||\mathcal{S}_{\lambda}(\bar{\rho}_A(\pi)) \bigr)~,
\end{equation}
where the average state is given by
\begin{equation}
    \bar{\rho}_A(\pi):= \int \ketbra{\alpha}{\alpha} p_N(\alpha) d^2\alpha~.
\end{equation}
If $H(BC)_{\mathcal{S}_{\lambda}( \bar{\rho}_A(\pi))} < \infty$, the Holevo quantity takes the simpler expression \cite{holevo2006continuous}
\begin{equation}
    \label{ap:gen_ens2}
    \chi_N\big(\mathcal{S}_{\lambda}(\pi)\big) = H(BC)_{\mathcal{S}_{\lambda}( \bar{\rho}_A(\pi))} - \int H(BC)_{\mathcal{S}_\lambda(\ketbra{\alpha}{\alpha})} p_N(\alpha) d^2 \alpha~.
\end{equation}
Recalling the $P$-representation of a thermal state \cite{glauber1963coherent}, i.e.,
\begin{equation}
    \label{eq:p-rep}
    \tau_{\bar{n}} = \frac{1}{\pi \bar{n}}\int d^2 \alpha ~e^{-|\alpha|^2/\bar{n}} \ketbra{\alpha}{\alpha} = \sum_{m=0}^{\infty} \frac{\bar{n}^m}{(1+ \bar{n})^{m+1}}\ketbra{m}{m}~,  
\end{equation}
we easily find $\bar{\rho}_A(\pi)=\tau_N$.
Let us now observe that, using Eq.~(\ref{eq:action_switch}), the expression of $\mathcal{S}_{\lambda}(\tau_N)$ in the qubit basis of $C$ reads
\begin{equation}
\label{eq:sigmaBC}
    \mathcal{S}_{\lambda}(\tau_N) = 
    \frac{1}{2}
    \begin{bmatrix}
    A  &  B\\
     B  & A
    \end{bmatrix}~,
\end{equation}
where we have set
\begin{align}
\label{eq:a}
    A &:= \int d^{2}\alpha ~p_N(\alpha)  \ketbra{\lambda \alpha}{\lambda \alpha}~,\\
    B &:= \int d^{2}\alpha ~p_N(\alpha)  e^{- \kappa_{\lambda} |\alpha|^2} \ketbra{\lambda \alpha}{\lambda \alpha}~.
\end{align}

\noindent Using Eq.~(\ref{eq:p-rep}), we find
\begin{equation}
    A = \tau_{\lambda^2N} = \sum_{n=0}^{\infty} a_n \ketbra{n}{n}~, \quad a_{n} := \frac{(N\lambda^{2})^n}{(1+N\lambda^2)^{n+1}}~.
\end{equation}
To evaluate $B$, we first observe that
\begin{equation}
    p_N(\alpha)e^{-\kappa_\lambda|\alpha|^2}
    =
    \frac{1}{1+\kappa_\lambda N}\,
    p_{N/(1+\kappa_\lambda N)}(\alpha)~,
\end{equation}
so that $B$ can be expressed as
\begin{equation}
    B = \frac{1}{1+\kappa_\lambda{N}}\int d^{2}\alpha ~p_{N/(1+\kappa_\lambda N)}(\alpha) \ketbra{\lambda \alpha}{\lambda \alpha}~.
\end{equation}

\noindent Making use of Eq.~(\ref{eq:p-rep}) therefore yields
\begin{equation}
    B =  \frac{1}{1+\kappa_\lambda N}\,
    \tau_{\lambda^2N/(1+\kappa_\lambda N)} = \sum_{n=0}^{\infty} b_n\ketbra{n}{n}~, \quad b_{n} := \frac{(N\lambda^{2})^n}{[1+(\lambda^2 +\kappa_\lambda)N]^{n+1}}~,
\end{equation}
from which Eq.~(\ref{eq:sigmaBC}) takes the form
\begin{equation}
    \mathcal{S}_{\lambda}(\tau_N) = \frac{1}{2}\sum_{n=0}^{\infty} 
 \ketbra{n}{n} \otimes \begin{bmatrix}
    a_n  &  b_n\\
     b_n & a_n
    \end{bmatrix}~.
\end{equation}
A straightforward calculation shows that the eigenvalues of $\mathcal{S}_{\lambda}(\tau_N)$ are given by $\{\nu_{n,\pm} := \frac{a_{n} \pm b_{n}}{2}\}$, for $n\in \mathbb{N}_{0}$.

\noindent Setting $c_{n} := b_{n}/a_{n}$, for $a_n>0$ we find
\begin{equation}
    c_{n} = \left[\frac{1+N\lambda^{2}}{1+(\lambda^2 +\kappa_\lambda)N}\right]^{n+1}~.
\end{equation}
Note that the eigenvalues $\nu_{n,\pm}$ can be expressed as $\nu_{n,\pm}= a_{n}\left(\frac{1 \pm c_{n}}{2}\right)$. Thus, we obtain
\begin{align}
    \label{eq:joint_entro_switch}
    H(BC)_{\mathcal{S}_{\lambda}(\tau_N)} = -\sum_{n=0}^{\infty}\sum_{s\in\{+,-\}}\nu_{n,s} \log_{2}(\nu_{n,s}) = - \sum_{n=0}^{\infty}a_{n} \log_{2}(a_{n}) +\sum_{n=0}^{\infty}a_{n}h_{2}\left(\frac{1+c_{n}}{2}\right)~,
\end{align}
where $h_2(p):=-(1-p)\log_2(1-p) -p \log_2(p)$ denotes the binary entropy function. 
Since $\{a_n\}_{n=0}^{\infty}$ is the probability distribution of a thermal state with mean photon number $N \lambda^2$, we immediately find (see, e.g., Eq.~(25.217) in \cite{wilde2017quantum})
\begin{equation}
H(\{a_n\}):=- \sum_{n=0}^{\infty}a_{n} \log_{2}(a_{n})= g(N \lambda^2)~,
\end{equation}
with $g(x) := (x+1) \log_2 (x+1) - x \log_2 (x)$. 
Thus, Eq.~(\ref{eq:joint_entro_switch}) reads
\begin{equation}
    H(BC)_{\mathcal{S}_{\lambda}(\tau_N)}= g(N\lambda^2) +\sum_{n=0}^{\infty}a_{n}h_{2}\left(\frac{1+c_{n}}{2}\right) \leq g(N\lambda^2) + 1~,
\end{equation}
where the last inequality follows from subadditivity, since $B$ is in a thermal state and $C$ is a qubit. 

\noindent Since $H(BC)_{\mathcal{S}_{\lambda}(\tau_N)}< \infty$ we can evaluate $\chi_{N}(\mathcal{S}_\lambda(\pi))$ using Eq.~(\ref{ap:gen_ens2}). We find
\begin{equation}
\label{eq:final_mutual}
    \chi_{N}(\mathcal{S}_{\lambda}(\pi)) =  g(N\lambda^2) +\Delta_N~,
\end{equation}
where we have defined
\begin{equation}
\label{eq:delta_c}
\Delta_N:=\sum_{n=0}^{\infty}a_{n}h_{2}\left(\frac{1+c_{n}}{2}\right) - \int d^2\alpha \frac{e^{-|\alpha|^2/N}}{\pi N} h_{2}\left(\frac{1+ e^{-\kappa_\lambda |\alpha|^2}}{2}\right)~.    
\end{equation}
\noindent Remarkably, $g(N\lambda^2)$ coincides with the energy-constrained classical capacity of a single-mode pure-loss channel with transmissivity $\lambda^2$ and mean input photon number $N$ \cite{giovannetti2004classical}, i.e.,  $\mathcal{C}_N(\mathcal{E}_{\lambda^2}) = g(N\lambda^2)$.
Thus, we can write
\begin{equation}
\label{eq:final_holevo}
    \chi_{N}(\mathcal{S}_{\lambda}(\pi)) =  \mathcal{C}_N(\mathcal{E}_{\lambda^2}) +\Delta_N~.
\end{equation}
We conclude by noting that
\begin{align}
    \sum_{n=0}^{\infty}a_{n}h_{2}\left(\frac{1+c_{n}}{2}\right) \leq \sum_{n=0}^{\infty}a_{n}=1~,
\end{align}
where we have used the fact that $0 \leq h_2(x) \leq 1$, for all $x \in [0,1]$. Similarly, we have
\begin{equation}
    \int d^2\alpha \frac{e^{-|\alpha|^2/N}}{\pi N} h_{2}\left(\frac{1+ e^{-\kappa_\lambda |\alpha|^2}}{2}\right) \le \int d^2\alpha \frac{e^{-|\alpha|^2/N}}{\pi N}=1~,
\end{equation}
since $\pi(d\alpha)$ is a probability measure on the complex space, thus ensuring $\Delta_N < \infty$.

\section{Proof of Eq.~(\ref{eq:res3})}
\label{ap:res3}
We show that $\mathcal{C}_{N}(\mathcal{S}_{\lambda}) > \mathcal{C}_{N}(\mathcal{E}_{\lambda^2})$, i.e., the classical capacity of a quantum switch with two pure-loss channels is strictly larger than that of a sequential use of such channels. 
We first prove that $\Delta_N$ in Eq.~(\ref{eq:final_mutual}) is strictly positive in $0 \leq \lambda < 1$.
To this end, let us first compute the Holevo quantity $\chi_N(\mathcal{E}_{\lambda^2}(\pi))$ given by
\begin{equation}
   \chi_N(\mathcal{E}_{\lambda^2}(\pi)) =  \int d^2\alpha ~p_N(\alpha) D\bigl(\mathcal{E}_{\lambda^2}(\rho_A)||\mathcal{E}_{\lambda^2}(\bar{\rho}_A(\pi)) \bigr)~.
\end{equation}
Setting $\rho_A= \ketbra{\alpha}{\alpha}$ and recalling that $\bar{\rho}_A(\pi)= \tau_N$ (see Appendix~\ref{ap:res2}), we obtain
\begin{equation}
    \mathcal{E}_{\lambda^2}(\tau_N) = \tau_{\lambda^2 N}~,
\end{equation}
where we have used the P-representation of Eq.~(\ref{eq:p-rep}) and the fact that $\mathcal{E}_{\lambda^2}(\ketbra{\alpha}{\alpha}) =\ketbra{\lambda \alpha}{\lambda \alpha} $.
As a consequence, $H(B)_{\mathcal{E}_{\lambda^2}(\tau_N)} = g(N \lambda^2) < \infty$, from which the Holevo quantity reads
\begin{equation}
    \chi_N \big(\mathcal{E}_{\lambda^2}(\pi)\big)  = H(B)_{\mathcal{E}_{\lambda^2}(\tau_N)} - \int d^{2}\alpha ~p_N(\alpha) H(B)_{\ketbra{\lambda \alpha}{\lambda \alpha}} = g(N \lambda^2)~,
\end{equation}
where the last equality follows from the fact that $H(B)_{\ketbra{\lambda \alpha}{\lambda \alpha}}=0$, since $B$ is in a pure state. 

\noindent As a consequence, the quantity $\Delta_N$ in Eq.~(\ref{eq:final_mutual}) can be cast as
\begin{equation}
    \Delta_N = \chi_N\big(\mathcal{S}_{\lambda}(\pi)\big)- \chi_N \big(\mathcal{E}_{\lambda^2}(\pi) \big)~.
\end{equation}
Introducing the states
\begin{align}
    &\sigma^{\alpha}_{BC}:=\mathcal{S}_{\lambda}(\ketbra{\alpha}{\alpha})~, \quad &&\bar{\sigma}_{BC}:= \mathcal{S}_{\lambda}(\tau_N)~,\\
    &\sigma^{\alpha}_{B}:=\mathcal{E}_{\lambda^2}(\ketbra{\alpha}{\alpha})~, \quad &&\bar{\sigma}_{B}:= \mathcal{E}_{\lambda^2}(\tau_N)~,
\end{align}
we can equivalently rewrite $\Delta_N$ as
\begin{align}
\Delta_N &=\int d^2 \alpha ~p_N(\alpha)\left[ D\bigl(\mathcal{S}_{\lambda}(\rho_A)||\mathcal{S}_{\lambda}(\bar{\rho}_A(\pi)) \bigr)- D\bigl(\mathcal{E}_{\lambda^2}(\rho_A)||\mathcal{E}_{\lambda^2}(\bar{\rho}_A(\pi)) \bigr) \right]\\
&=\int d^2 \alpha ~p_N(\alpha)\left[ D(\sigma^{\alpha}_{BC}||\bar{\sigma}_{BC})- D(\sigma^{\alpha}_{B}||\bar{\sigma}_{B}) \right]\\
\label{eq:dpi}
&=\int d^2 \alpha ~p_N(\alpha)\big[ D(\sigma^{\alpha}_{BC}||\bar{\sigma}_{BC})- D\big(\mbox{Tr}_{C}(\sigma^{\alpha}_{BC})||\mbox{Tr}_{C}(\bar{\sigma}_{BC})\big) \big]\geq 0~,
\end{align}
where we have used data processing inequality.
Now let us assume $\Delta_N=0$, which implies $ D(\sigma^{\alpha}_{BC}||\bar{\sigma}_{BC})= D(\sigma^{\alpha}_{B}||\bar{\sigma}_{B})$ for $\pi$-almost every $\alpha$. Let $\mathcal A_0\subseteq\mathbb C$ be the set on which the relative entropies are finite and this
equality holds. By Theorem 2 in \cite{jenvcova2024recoverability}, there exists a Petz recovery channel $\mathcal{R}:=\mathcal{R}_{B\rightarrow BC}$, associated with $\mbox{Tr}_C$ and the reference state $\bar{\sigma}_{BC}$, that satisfies
\begin{equation}
\label{eq:recovery}
        \mathcal R(\sigma_B^\alpha)=\sigma_{BC}^\alpha,
    \qquad
    \mathcal R(\bar\sigma_B)=\bar\sigma_{BC},
\end{equation}
for every $\alpha \in \mathcal{A}_0$. Since the reference state $\bar{\sigma}_{BC}$ and the channel $\operatorname{Tr}_C$ are independent of $\alpha$, the same recovery map can be chosen for all amplitudes in $\mathcal{A}_0$.
Choose $\alpha, \beta \in \mathcal{A}_0$ with $|\alpha|^2\neq|\beta|^2$. 
Eq.~(\ref{eq:action_switch}) gives
\begin{equation}
    \sigma_{BC}^\alpha
    =
    \sigma_B^\alpha \otimes \sigma_C^\alpha = \ketbra{\lambda \alpha}{\lambda \alpha} \otimes \sigma_C^\alpha~.
\end{equation}
Since $\kappa_\lambda>0$ for $0\leq\lambda<1$, the different magnitudes imply
$\sigma_C^\alpha \neq \sigma_C^\beta$.

Using the squared fidelity $F(\rho,\sigma):=\left(\mbox{Tr}\sqrt{\sqrt{\rho} \sigma \sqrt{\rho}} \right)^2$, define
\begin{equation}
    f_{\alpha\beta}
    :=
    F(\sigma_B^\alpha,\sigma_B^\beta)
    =
    |\braket{\lambda\alpha|\lambda \beta}|^2 
    >0~.
\end{equation}
Monotonicity of fidelity under quantum channels and multiplicativity on tensor products \cite{wilde2017quantum} imply
\begin{equation}
\begin{aligned}
    f_{\alpha\beta}
    &\leq
    F\!\left(
        \mathcal R(\sigma_B^\alpha),
        \mathcal R(\sigma_B^\beta)
    \right)
    \\
    &=
    F(\sigma_{BC}^\alpha,\sigma_{BC}^\beta)
    \\
    &=
    f_{\alpha\beta}
    F(\sigma_C^\alpha,\sigma_C^\beta)~.
\end{aligned}
\end{equation}
Dividing both sides by $f_{\alpha\beta}>0$, we find $ F(\sigma_C^\alpha,\sigma_C^\beta) \geq 1$, which implies $F(\sigma_C^\alpha,\sigma_C^\beta)=1$, contradicting the initial assumption of distinct states. This proves that $\Delta_N >0$. Hence,
\begin{equation}
\label{eq:strict_i}
    \chi_{N}(\mathcal{S}_{\lambda}(\pi)) = \mathcal{C}_{N}(\mathcal{E}_{\lambda^2}) + \Delta_N >  \mathcal{C}_{N}(\mathcal{E}_{\lambda^2})~,
\end{equation}
where we have used the fact that $\mathcal{C}_{N}(\mathcal{E}_{\lambda^2}) = g(N \lambda^2)$. To extend the comparison to the energy-constrained classical capacity, consider $n$ independent draws from the input ensemble, denoted by $\pi^{\otimes n}$.
Its average state is $\tau_N^{\otimes n}$ and satisfies
\begin{equation}
    \operatorname{Tr}\big[
        \big(\sum_{j=1}^{n} \hat{n}_{A_j}\big)\tau_N^{\otimes n}\big]= nN~.
\end{equation}
Additivity of quantum relative entropy on product states gives
\begin{equation}
\chi_{nN} \left(
\mathcal S_\lambda^{\otimes n}(\pi^{\otimes n})
\right) = n \chi_{N}\big (\mathcal{S}_\lambda(\pi) \big).
\end{equation}
Dividing by $n$ and using the regularization formula
given in the main text \cite{holevo2003entanglement},
we obtain
\begin{equation}
    \mathcal{C}_N(\mathcal{S}_\lambda) \geq  \chi_N(\mathcal{S}_\lambda(\pi))> \mathcal{C}_N(\mathcal{E}_{\lambda^2})~,
\end{equation}
which proves Eq.~(\ref{eq:res3}).

\section{Proof of Eq.~(\ref{eq:res4})}
\label{ap:res4}
We show that $\mathcal{Q}_{N}(\mathcal{S}_{\lambda}) > \mathcal{Q}_{N}(\mathcal{E}_{\lambda^2})$, i.e., the switch has a strictly higher quantum capacity than the sequential channel. We consider $0<N<\infty$ and first derive the explicit expression of $\Delta_N^Q$ in Eq.~(\ref{eq:res_coherent}) of the main text.
In what follows, it will be convenient to use an alternative formulation for $I_c(\rho, \mathcal{L})$ given by \cite{holevo2010mutual, wilde2018energy}
\begin{equation}
\label{eq:coh_info}
    I_c(\rho, \mathcal{L}) = H(B)_{\omega}-H(RB)_{\omega}~,
\end{equation}
where $\omega:=\omega_{RB}=(\mathds{1}_{R} \otimes \mathcal{L})(\ketbra{\Psi}{\Psi})$,  $\ket{\Psi} \in \mathcal{H}_{R} \otimes \mathcal{H}_{A}$ is a purification of $\rho$ with $\mathcal{H}_{R} \simeq \mathcal{H}_{A}$ and all the entropies are finite. 

Let us consider the thermal state $\tau_N$ given by
\begin{equation}
\label{eq:thermal}
    \tau_N =\sum_{n=0}^{\infty}t_n\ketbra{n}{n},
    \qquad
    t_n:=\frac{N^n}{(N+1)^{n+1}},
\end{equation}
and choose its purification to be the two-mode squeezed vacuum state
\begin{equation}
    \ket{\Psi_N}  = \sum_{n=0}^{\infty}\sqrt{t_n}\ket{n}_R\ket{n}_A.
\end{equation}
The joint reference-output state of the switch is
\begin{equation}
\label{eq:evol_puri}
    \omega_{RBC}
    :=
    (\operatorname{id}_R\otimes\mathcal S_\lambda)
    (\ketbra{\Psi_N}{\Psi_N}).
\end{equation}
Recalling that $\mbox{Tr}_{C} \circ \mathcal{S}_{\lambda} = \mathcal{E}_{\lambda^2}$,  it is easy to see that $\omega_{RBC}$ satisfies
\begin{equation}
\label{eq:pur}
    \omega_{RB}:=\mbox{Tr}_{C}(\omega_{RBC})=:\zeta_{RB}~,
\end{equation}
where we have set $\zeta_{RB}:=(\mathds{1}_{R} \otimes \mathcal{E}_{\lambda^2})(\ketbra{\Psi_N}{\Psi_N})$.
Note that all entropies are finite since $H(R)_\omega=g(N)$, while Appendix~\ref{ap:res2} gives $H(BC)_\omega\leq g(\lambda^2N)+1$.
Subadditivity therefore implies
\begin{equation}
    H(RBC)_\omega
    \leq g(N)+g(\lambda^2N)+1<\infty.
\end{equation}
The remaining marginal entropies are finite as well. 
Hence, using Eq.~(\ref{eq:coh_info}) we find
\begin{align}
    &I_c(\tau_N, \mathcal{S}_{\lambda})=H(BC)_\omega - H(RBC)_\omega~,\\
    &I_c(\tau_N, \mathcal{E}_{\lambda^2}) = H(B)_\zeta - H(RB)_\zeta~.
\end{align}
Introducing 
\begin{equation}
    \Delta_N^{Q}:=I_c(\tau_N, \mathcal{S}_{\lambda})-I_c(\tau_N, \mathcal{E}_{\lambda^2})~,
\end{equation}
we easily obtain
\begin{equation}
\label{eq:delta_q}
    \Delta_N^{Q}=H(BC)_\omega - H(RBC)_\omega-H(B)_\omega + H(RB)_\omega~,
\end{equation}
from which direct calculation yields
\begin{align}
    \Delta_N^{Q} &= H(BC)_\omega - H(RBC)_\omega-H(B)_\omega + H(RB)_\omega \\
    &= H(R|B)_\omega + H(C|B)_\omega-H(RC|B)_\omega = I(R;C|B)_\omega~,
\end{align}
where we have used the definition of the conditional quantum mutual information $I(R;C|B)_\omega$. Hence,
\begin{equation}
\label{eq:qc_enhancement}
    I_c(\tau_N, \mathcal{S}_{\lambda}) = I_c(\tau_N, \mathcal{E}_{\lambda^2})+ \Delta_N^Q~.
\end{equation}
Strong subadditivity guarantees $\Delta_N^Q\geq0$. To establish strict positivity, we proceed by contradiction, as in Appendix~\ref{ap:res3}, and assume $\Delta_N^Q=0$. Proposition 13 in \cite{shirokov2016measures} implies the existence of a recovery map $\mathcal{R}:=\mathcal{R}_{B \rightarrow BC}$, such that $\omega_{RBC}=(\mathds{1}_{R}\otimes \mathcal{R})(\omega_{RB})$, whose action is given by
\begin{equation}
   (\mathds{1}_{R} \otimes \mathcal{R} \circ \mathcal{E}_{\lambda^2})(\ketbra{\Psi_{N}}{\Psi_{N}}) = (\mathds{1}_{R} \otimes \mathcal{S}_{\lambda})(\ketbra{\Psi_{N}}{\Psi_{N}})~.
\end{equation}
Expanding in the Fock basis and taking diagonal elements on the reference system gives
\begin{equation}
    \mathcal{R} \circ \mathcal{E}_{\lambda^2}(\ketbra{n}{n}) = \mathcal{S}_{\lambda} (\ketbra{n}{n})~,
\end{equation}
where we have used the fact that $t_{n} > 0$ for all $n$.

Let us now consider the squared fidelity $F(\rho,\sigma):=\left(\mbox{Tr}\sqrt{\sqrt{\rho} \sigma \sqrt{\rho}} \right)^2$ between two quantum states $\rho_i := \mathcal{E}_{\lambda^2}(\ketbra{i}{i})$ for $i \in \{0,1\}$. Since $\mathcal{R}$ is a quantum channel, monotonicity of the fidelity under CPTP maps implies
\begin{equation}
\label{eq:fidelity_comp}
    F(\rho_0, \rho_1) \leq F(\mathcal{R}(\rho_0), \mathcal{R}(\rho_1))~.
\end{equation}
Recalling the expression of the Kraus operators of Eq.~(\ref{eq:kraus_double}), direct calculation yields
\begin{align}
    &\rho_0= \ketbra{0}{0}~,\\
    &\rho_1= \lambda^2 \ketbra{1}{1} + (1-\lambda^2)\ketbra{0}{0}~.
\end{align}
Conversely, using Eq.~(\ref{eq:switch_mat_5}) we find
\begin{align}
    &\mathcal{R}(\rho_0)= \ketbra{0}{0} \otimes \sigma_0~,\\
    &\mathcal{R}(\rho_1)= \lambda^2\ketbra{1}{1} \otimes \sigma_0 + (1-\lambda^2~)\ketbra{0}{0} \otimes \sigma_1~.
\end{align}
Diagonalization of $\sigma_0, \sigma_1$ yields
\begin{align}
    &\sigma_0 = \ketbra{+}{+}~,\\
    &\sigma_1 =\left(\frac{1+\gamma}{2}\right)\ketbra{+}{+}+ \left(\frac{1-\gamma}{2}\right)\ketbra{-}{-}~.
\end{align}
Hence, we can write
\begin{equation}
    F(\rho_0,\rho_1) = \braket{0|\rho_1|0} = 1-\lambda^2~,
\end{equation}
where we have used the fact that $\rho_0$ is a pure state.
Similarly, we have
\begin{equation}
    F(\mathcal{R}(\rho_0),\mathcal{R}(\rho_1)) = \braket{0,+|\mathcal{R}(\rho_1)|0,+} = \frac{1}{2}(1-\lambda^2)(1+\gamma)~.
\end{equation}
A direct comparison shows that Eq.~(\ref{eq:fidelity_comp}) holds true only if $(1+\gamma)/2\geq1$, which requires $\lambda \geq 1$. Since $\lambda \in[0,1]$, this is possible only at $\lambda=1$. Thus, for $0\leq\lambda<1$, we obtain a contradiction, establishing that $\Delta_N^Q>0$. 

The achievability of the coherent information in the present infinite-dimensional setting follows from Theorem 5 of Wilde and Qi~\cite{wilde2018energy}, which establishes a regularized formula for the energy-constrained quantum capacity under two requirements: (i) the input energy observable is a Gibbs observable, and (ii) the channel satisfies the finite-output-entropy condition
\begin{equation}
\sup_{\rho_A:\,\operatorname{Tr}(\rho_A\hat n_A)\leq N}
H\big(BC\big)_{\mathcal{S}_\lambda(\rho_A)}<\infty~.
\end{equation}
Both conditions are satisfied in our setting. Indeed, in our case, the input energy observable is given by the photon-number operator $\hat n_A$, which obeys the Gibbs condition
\begin{equation}
\operatorname{Tr}(e^{-\beta\hat n_A})
=\frac{1}{1-e^{-\beta}}<\infty
\end{equation}
for every $\beta>0$. As for the second condition, note that a pure-loss channel $\mathcal{E}_{\lambda}$ reduces the mean
input photon number by a factor $\lambda$ \cite{giovannetti2004classical}.
Thus, for any input state $\rho_A$ satisfying
$\operatorname{Tr}[\hat n_A\rho_A]\leq N$, let
\begin{equation}
    \sigma_{BC}:=\mathcal S_\lambda(\rho_A),
    \qquad
    \sigma_B:=\operatorname{Tr}_C(\sigma_{BC}),
    \qquad
    \sigma_C:=\operatorname{Tr}_B(\sigma_{BC}).
\end{equation}
The identity
$\operatorname{Tr}_C\circ\mathcal S_\lambda
=\mathcal E_{\lambda^2}$ then gives
\begin{equation}
    \operatorname{Tr}(\sigma_B\hat n_B)
    =\lambda^2\operatorname{Tr}(\rho_A\hat n_A)
    \leq\lambda^2N~.
\end{equation}
Since the thermal state maximizes the von Neumann entropy among single-mode bosonic states with a given mean photon number, subadditivity establishes
\begin{equation}
H(BC)_{\sigma_{BC}}
\leq H(B)_{\sigma_B}+H(C)_{\sigma_C}
\leq g(\lambda^2N)+1<\infty~,
\end{equation}
where we have used the fact that $H(C)_{\sigma_C} \leq1$, since $C$ is a qubit. As a consequence, since $\tau_N^{\otimes k}$ is an admissible input state, Theorem 5 in \cite{wilde2018energy}, together with the additivity of coherent information on tensor-product states, guarantees that $I_c(\tau_N,\mathcal S_\lambda)$ is an achievable asymptotic quantum communication rate. For $N>0$ and $\frac{1}{\sqrt{2}}\leq\lambda<1$, we therefore obtain
\begin{equation}
\mathcal{Q}_N(\mathcal{S}_\lambda)
\geq I_c(\tau_N,\mathcal{S}_\lambda)
>\mathcal{Q}_N(\mathcal{E}_{\lambda^2})~.
\end{equation}

\section{Proof of Eq.~(\ref{eq:res5})}
\label{ap:res5}

We show that the quantum switch can activate the energy-constrained quantum capacity of a pure-loss channel in a region where this is zero. To establish this result, we restrict the input to the subspace spanned by the vacuum and single-photon states, obtaining an effective qubit channel whose quantum capacity lower-bounds that of the original switch. We then derive its Kraus representation and construct its complementary channel. By explicitly constructing a degrading map over a suitable parameter range, we reduce the energy-constrained quantum capacity of the restricted channel to a single-letter optimization. Finally, we identify an admissible input state with strictly positive coherent information. Comparing the resulting lower bound with the energy-constrained quantum capacity of the sequential pure-loss channel $\mathcal E_{\lambda^2}$ reveals a regime in which the switch supports reliable quantum communication while the corresponding definite-order composition has zero quantum capacity. For the ease of readability, we split the proof in three steps.

\subsection{The restricted qubit channel}
Let $\mathcal{S}^{(2)}_{\lambda}$ denote the restriction of
$\mathcal{S}_{\lambda}$ to the qubit subspace $\mathcal{H}^{(2)}:=\mbox{span}\{\ket{0},\ket{1}\}$. 
On this subspace, the photon-number operator is given by $\hat n_A^{(2)}=\ketbra{1}{1}$. Thus, the energy constraint becomes
$\operatorname{Tr}(\hat{n}^{(2)}_A\rho_A)=\braket{1|\rho_A|1} \leq N$. Any code for the restricted channel can be embedded into the bosonic input space without changing its input photon number.
Consequently, $\mathcal{Q}_N(\mathcal{S}_{\lambda})
    \geq
    \mathcal{Q}_N(\mathcal{S}^{(2)}_{\lambda})$. Restricting Eq.~(\ref{eq:switch_mat_5}) to $\mathcal{H}^{(2)}$, we find
\begin{align}
 \mathcal{S}^{(2)}_\lambda(\rho_A) &= 
    \sum_{m=0}^{1} K_{m}^{\lambda^2}\rho_A \left(K_{m}^{\lambda^2} \right)^\dagger \otimes
    \sigma_{m}\\
    &= K_{0}^{\lambda^2}\rho_A \left(K_{0}^{\lambda^2} \right)^\dagger \otimes
    \sigma_{0} + K_{1}^{\lambda^2}\rho_A \left(K_{1}^{\lambda^2} \right)^\dagger \otimes
    \sigma_{1}\\
    &= K_{0}^{\lambda^2}\rho_A \left(K_{0}^{\lambda^2} \right)^\dagger \otimes
    \ketbra{+}{+} + K_{1}^{\lambda^2}\rho_A \left(K_{1}^{\lambda^2} \right)^\dagger \otimes
    \left[\frac{1+\gamma}{2}\ketbra{+}{+}+ \frac{1-\gamma}{2}\ketbra{-}{-}\right]~,
\end{align}
where we have diagonalized the matrices $\sigma_0, \sigma_1$ of Eq.~(\ref{eq:sigma}) and introduced the states $\ket{\pm} := \frac{\ket{0} \pm \ket{1}}{\sqrt{2}}$. Here,
\begin{align}
    &K_{0}^{\lambda^2}=\ketbra{0}{0} + \lambda \ketbra{1}{1}~,\\
    &K_{1}^{\lambda^2}=\sqrt{1-\lambda^2} \ketbra{0}{1}~,
\end{align}
from which $\mathcal{S}^{(2)}_\lambda$ admits the Kraus operators
\begin{align}
\label{eq:kraus_res_1}
    &L_0:= \left(\ketbra{0}{0} + \lambda \ketbra{1}{1} \right) \otimes \ket{+} = \ketbra{0,+}{0} + \lambda \ketbra{1,+}{1}~,\\
\label{eq:kraus_res_2}
    &L_1:= \sqrt{\lambda_{+}} \ketbra{0}{1} \otimes \ket{+} = \sqrt{\lambda_{+}} \ketbra{0,+}{1}~,\\
\label{eq:kraus_res_3}
    &L_2:= \sqrt{\lambda_{-}} \ketbra{0}{1} \otimes \ket{-} = \sqrt{\lambda_{-}} \ketbra{0,-}{1}~,
\end{align}
where we have set $\lambda_{\pm}:=\frac{(1\pm\gamma)(1-\lambda^2)}{2}$.
In particular, $\lambda^2+\lambda_++\lambda_-=1$, which ensures
$\sum_{i=0}^2 L_{i}^\dagger L_{i}=\mathds{1}_A$.
Assuming an input state $\rho_A$ of the form
\begin{equation}
    \rho_A =
    \begin{bmatrix}
        \rho_{00} & \rho_{01}\\
        \rho_{01}^{*} & \rho_{11}
    \end{bmatrix}~,
\end{equation}
the representation of the restricted channel $\mathcal{S}^{(2)}_\lambda$ in the basis $\{\ket{0,+},\ket{1,+},\ket{0,-}\}$ reads
\begin{equation}
\label{eq:res_channel}
    \mathcal{S}^{(2)}_\lambda(\rho_A)=
     \begin{bmatrix}
        \rho_{00} + \lambda_{+} \rho_{11} & \lambda \rho_{01} & 0 \\
        \lambda \rho_{01}^{*} & \lambda^2\rho_{11} & 0 \\
        0 & 0 &\lambda_{-}\rho_{11}
    \end{bmatrix}~.
\end{equation}

\subsection{Degradability of the restricted channel}
Let us consider the isometric extension 
\begin{equation}
    V= \sum_{i=0}^{2}L_{i}\otimes \ket{e_{i}}_E~,
\end{equation}
where $\{\ket{e_{i}}\}_{i=0}^{2}$ is an orthonormal basis for the environment $E$. Then, the complementary channel $\hat{\mathcal{S}}^{(2)}_{\lambda}$ from the restricted subspace of $A$ to $E$ is given by
\begin{equation}
\label{eq:compl}
    \hat{\mathcal{S}}^{(2)}_{\lambda}(\rho_A) :=\mbox{Tr}_{BC}(V \rho_A V^\dagger) = \sum_{i,j=0}^{2}\mbox{Tr}[L_i \rho_A L_{j}^{\dagger}]\ketbra{e_i}{e_j}~.
\end{equation}
An explicit calculation of the diagonal terms in Eq.~(\ref{eq:compl}) yields
\begin{align}
    &\mbox{Tr}[L_0 \rho_A L_{0}^{\dagger}] = \rho_{00} + \lambda^2 \rho_{11} ~,\\
    &\mbox{Tr}[L_1 \rho_A L_{1}^{\dagger}] = \lambda_{+}\rho_{11}~,\\
    &\mbox{Tr}[L_2 \rho_A L_{2}^{\dagger}] = \lambda_{-}\rho_{11}~,
\end{align}
while the only non-zero off-diagonal term is
\begin{align}
    &\mbox{Tr}[L_0 \rho_A L_{1}^{\dagger}] = \sqrt{\lambda_{+}} \rho_{01}~,
\end{align}
together with its corresponding Hermitian conjugate.
As a consequence, the representation of the complementary channel $\hat{\mathcal{S}}^{(2)}_{\lambda}(\rho_A)$ in the environmental basis $\{\ket{e_{i}}\}_{i=0}^{2}$ reads
\begin{equation}
\label{eq:comp_channel}
    \hat{\mathcal{S}}^{(2)}_{\lambda}(\rho_A)= 
    \begin{bmatrix}
        \rho_{00} + \lambda^2 \rho_{11} & \sqrt{\lambda_{+}} \rho_{01} & 0 \\
        \sqrt{\lambda_{+}} \rho_{01}^{*} & \lambda_{+}\rho_{11} & 0 \\
        0 & 0 &\lambda_{-}\rho_{11}
    \end{bmatrix}~.
\end{equation}
A direct comparison reveals that Eq.~(\ref{eq:res_channel}) and Eq.~(\ref{eq:comp_channel}) have the same expression upon the transformation $\lambda \longleftrightarrow \sqrt{\lambda_{+}}$. This observation suggests the existence of a degrading map $\mathcal{D}$, such that $\hat{\mathcal{S}}^{(2)}_{\lambda} = \mathcal{D} \circ \mathcal{S}^{(2)}_{\lambda}$, with Kraus operators 
\begin{align}
    &D_{0}:=\ketbra{e_0}{b_0} + \sqrt{\eta} \ketbra{e_1}{b_1} + \ketbra{e_2}{b_2}~,\\
    &D_{1}:= \sqrt{1- \eta}  \ketbra{e_0}{b_1}~,
\end{align}
where $0 \leq \eta \leq 1$ and we have set $\ket{b_0}:=\ket{0,+}, \ket{b_1}:=\ket{1,+}, \ket{b_2}:=\ket{0,-}$.
These operators satisfy
\begin{equation}
    D_0^\dagger D_0+D_1^\dagger D_1
    = \sum_{i=0}^{2} \ketbra{b_i}{b_i}~,
\end{equation}
so that $\mathcal{D}$ is a CPTP map on the output space of the restricted channel.
Recalling Eq.~(\ref{eq:res_channel}), the action of $\mathcal{D}$ on the restricted channel yields
\begin{equation}
    \mathcal{D} \circ  \mathcal{S}^{(2)}_{\lambda}(\rho_A) = 
    \begin{bmatrix}
        \rho_{00} + [\lambda_{+}+(1-\eta)\lambda^2] \rho_{11} & \sqrt{\eta}\lambda \rho_{01} & 0 \\
        \sqrt{\eta}\lambda \rho_{01}^{*} & \eta \lambda^2\rho_{11} & 0 \\
        0 & 0 &\lambda_{-}\rho_{11}
    \end{bmatrix}~.
\end{equation}
For $\lambda>0$, choosing $\eta =\lambda_{+}/\lambda^2$ yields
\begin{equation}
    \hat{\mathcal{S}}^{(2)}_{\lambda}
    =\mathcal{D}\circ\mathcal{S}^{(2)}_\lambda~,
\end{equation}
thus establishing degradability of the restricted channel whenever $\lambda^2 \geq \lambda_+$. Remarkably, by invoking the contractivity of the trace distance, we can prove that this condition is also necessary. In fact, let us recall the definition of trace distance between two quantum states $\sigma_1,\sigma_2$, i.e.,
\begin{equation}
T(\sigma_1,\sigma_2):=\frac{1}{2}\norm{\sigma_1 - \sigma_2}_1~,
\end{equation}
where $\norm{\cdot}_1$ denotes the trace norm of an operator.
Then, for any CPTP map $\Phi$ it holds
\begin{equation}
    T(\Phi(\sigma_1),\Phi(\sigma_2)) \leq T(\sigma_1,\sigma_2)~.
\end{equation}
Let us assume that $\mathcal{S}^{(2)}_{\lambda}$ is degradable, so that $\Phi = \mathcal{D}$ is a CPTP map. Taking $\sigma_i= \mathcal{S}^{(2)}_{\lambda}(\rho_i)$ we find
\begin{equation}
    T(\mathcal{D}\circ\mathcal{S}^{(2)}_{\lambda}(\rho_1),\mathcal{D}\circ\mathcal{S}^{(2)}_{\lambda}(\rho_2)) \leq T(\mathcal{S}^{(2)}_{\lambda}(\rho_1),\mathcal{S}^{(2)}_{\lambda}(\rho_2))~,
\end{equation}
which implies
\begin{equation}
\label{eq:trace_norm_comp}
    \frac{1}{2}\norm{\hat{\mathcal{S}}^{(2)}_{\lambda}(\rho_1)-\hat{\mathcal{S}}^{(2)}_{\lambda}(\rho_2)}_1  \leq \frac{1}{2}\norm{\mathcal{S}^{(2)}_{\lambda}(\rho_1)-\mathcal{S}^{(2)}_{\lambda}(\rho_2)}_1~.
\end{equation}

\noindent Setting $\rho_i=\ketbra{\pm}{\pm}$ with $i\in \{1,2\}$, we find $T\left(\mathcal{S}^{(2)}_{\lambda}(\rho_1),\mathcal{S}^{(2)}_{\lambda}(\rho_2)\right) = \lambda$. A similar calculation for the complementary channel gives $T\left(\hat{\mathcal{S}}^{(2)}_{\lambda}(\rho_1),\hat{\mathcal{S}}^{(2)}_{\lambda}(\rho_2)\right)  =  \sqrt{\lambda_+}$ so that Eq.~(\ref{eq:trace_norm_comp}) becomes $\lambda^2 \geq \lambda_+$. We conclude that $\mathcal{S}^{(2)}_\lambda$ is degradable
if and only if $\lambda^2 \geq \lambda_{+}$.
Using
\begin{equation}
    \lambda_{+}
    =\frac{(1-\lambda)(1+\sqrt{\lambda})^2}{2},
\end{equation}
this condition becomes $\lambda \geq \lambda_*$, where $\lambda_{*} \approx 0.705776$. 

\subsection{Bounding the quantum capacity of the restricted channel}

Here, we provide an input state such that $I_c(\rho_A, \mathcal{S}^{(2)}_\lambda)>0$, which establishes an admissible lower bound to the quantum capacity $\mathcal{Q}_N(\mathcal{S}^{(2)}_{\lambda})$. 
Note that, since for $\lambda \geq \lambda_*$ the restricted channel is degradable, its quantum capacity can be expressed through the single-letter formula \cite{wilde2018energy}
\begin{equation}
    \mathcal{Q}_N(\mathcal{S}^{(2)}_\lambda)
    =
    \max_{\rho_A:~ \braket{1|\rho_A|1}\leq N}
    I_c(\rho_A,\mathcal{S}^{(2)}_\lambda)
\end{equation}
where the coherent information of $\mathcal{S}^{(2)}_\lambda$ is given by
\begin{equation}
    I_c(\rho_A, \mathcal{S}^{(2)}_\lambda):= H(BC)_{\mathcal{S}^{(2)}_\lambda(\rho_A)}-H(E)_{\hat{\mathcal{S}}^{(2)}_\lambda(\rho_A)}~.
\end{equation}

Let us consider the following quantum state:
\begin{equation}
    \rho_{p} = (1-p)\ketbra{0}{0} + p \ketbra{1}{1}~, \quad 0 \leq p \leq 1~. 
\end{equation}
Direct calculation shows that both output states are diagonal, i.e.,
\begin{align}
    &\mathcal{S}^{(2)}_\lambda(\rho_p)=
     \begin{bmatrix}
        1-p + \lambda_{+} p & 0 & 0 \\
       0 & \lambda^2 p & 0 \\
        0 & 0 &\lambda_{-}p
    \end{bmatrix}~,\\
        &\hat{\mathcal{S}}^{(2)}_{\lambda}(\rho_p)=
    \begin{bmatrix}
            1-p + \lambda^2 p & 0 & 0 \\
            0 & \lambda_{+}p & 0 \\
            0 & 0 &\lambda_{-}p
        \end{bmatrix}~.
\end{align}
Note that, due to the chain rule, the entropy of a quantum state with three eigenvalues $\{r_i\}_{i=1}^{3}$ can be written as
\begin{equation}
    H(r_1,r_2,r_3) = h_2(r_3) +(1-r_3)h_2\left(\frac{r_2}{1-r_3}\right)~,
\end{equation}
with $\sum_{i=1}^{3} r_i = 1$. Using the above formula we obtain
\begin{align}
    & H(\mathcal{S}^{(2)}_\lambda(\rho_p)) = h_2(\lambda_{-}p) + (1-\lambda_{-}p) h_2\left(\frac{\lambda^2 p}{1-\lambda_{-}p}\right)~,\\
     & H(\hat{\mathcal{S}}^{(2)}_{\lambda}(\rho_p)) = h_2(\lambda_{-}p) + (1-\lambda_{-}p) h_2\left(\frac{\lambda_{+}p}{1-\lambda_{-}p}\right)~.
\end{align}
Then, the coherent information reads
\begin{equation}
    I_c(\rho_p, \mathcal{S}^{(2)}_{\lambda}) =  (1-\lambda_{-}p) \left[h_2\left(\frac{\lambda^2 p}{1-\lambda_{-}p}\right) -h_2\left(\frac{\lambda_{+}p}{1-\lambda_{-}p}\right)\right]~.
\end{equation}
Let us now set $x:=\frac{\lambda^2p}{1-\lambda_{-}p}$ and $y:=\frac{\lambda_{+}p}{1-\lambda_{-}p}$.
For $0 < p < 1$, it is easy to see that
\begin{align}
    x + y=\frac{(\lambda^2 +\lambda_{+})p}{1-\lambda_{-}p} =\frac{(1 -\lambda_{-})p}{1-\lambda_{-}p} <1~,
\end{align}
where we have used the fact that $\lambda^2 + \lambda_{-} + \lambda_{+} =1$.
Moreover, $x > y \iff \lambda^2 > \lambda_{+}$.

Let us now assume that $x>y$. Since $x+y<1$, we also have $y<1-x$, implying $y<\min\{x,1-x\}\leq 1/2$. It follows that
\begin{equation}
h_2(y)<h_2\left(\min\{x,1-x\}\right)=h_2(x)~,
\end{equation}
where we have used $h_2(x) = h_2(1-x)$ and the fact that $h_2$ is strictly increasing on $[0,1/2]$. Conversely, if $x<y$, the same argument with $x$ and $y$ exchanged yields $h_2(x)<h_2(y)$. Hence, we have
\begin{equation}
    h_2(y)<h_2(x) \iff y < x~.
\end{equation}
Recalling the definitions of $x$ and $y$, we conclude that,
for every $0<p<1$,
\begin{equation}
    I_c(\rho_p,\mathcal{S}^{(2)}_\lambda)>0
    \iff \lambda^2>\lambda_{+}
    \iff \lambda>\lambda_{*},
\end{equation}
where $\lambda_{*} \approx 0.705776$.
For every $N>0$, we can choose $0 < p < \min \{N,1\}$, so that
$\operatorname{Tr}(\hat{n}_A^{(2)}\rho_{p})=p<N$.
Hence, for $\lambda>\lambda_{*}$,
\begin{equation}
    \mathcal{Q}_N(\mathcal{S}_\lambda)
    \geq \mathcal{Q}_N(\mathcal{S}_\lambda^{(2)})
    \geq I_c(\rho_{p},\mathcal{S}_\lambda^{(2)}) >0.
\end{equation}
Since the sequential channel has zero quantum capacity
for $\lambda \leq 1/\sqrt{2}$, we obtain
\begin{equation}
    \mathcal{Q}_N(\mathcal{S}_\lambda) >0
    =\mathcal{Q}_N(\mathcal{E}_{\lambda^2}),
    \quad
    \lambda_{*} < \lambda \leq \frac{1}{\sqrt{2}},
\end{equation}
which proves Eq.~(\ref{eq:res5}).

\section{Proof of Eq.~(\ref{eq:res6})}
\label{ap:res6}
We prove that, even when two-way classical communication is available, the quantum switch has strictly greater quantum capacity than the sequential channel. 
Let us consider again the thermal state $\tau_N$ of Eq.~(\ref{eq:thermal}) together with its purification, given by the TMSV state $\ket{\Psi_N}$, i.e.,
\begin{equation}
    \ket{\Psi_N}:=\sum_{n=0}^{\infty} \sqrt{t_n} \ket{n}_{R}\ket{n}_{A}~, \quad t_n := \frac{N^n}{(N+1)^{n+1}}~.
\end{equation}
Direct calculation reveals that the state $\omega_{RBC}$ of Eq.~(\ref{eq:evol_puri}) can be written as
\begin{align}
     \omega_{RBC} := (\mathds{1}_{R} \otimes \mathcal{S}_{\lambda})(\ketbra{\Psi_N}{\Psi_N}) = \sum_{m=0}^{\infty}\ketbra{\Phi_m}{\Phi_m}\otimes \sigma_m~,
\end{align}
where we have introduced the (unnormalized) states $\ket{\Phi_m}$ given by
\begin{equation}
    \ket{\Phi_m} :=(\mathds{1}_{R}\otimes K^{\lambda^2}_m)\ket{\Psi_{N}}~.
\end{equation}
Note that $\omega_R=\tau_N$ and
$\omega_B=\tau_{\lambda^2N}$. Thus, the entropy bounds established in Appendix~\ref{ap:res2},
together with $H(R)_\omega=g(N)<\infty$ and the fact that
$C$ is a qubit imply, by subadditivity, that all joint and
marginal entropies are finite.

Diagonalizing the states $\sigma_m$ in the basis $\{\ket{\pm}\}$ yields
\begin{equation}
    \omega_{RBC} = \sum_{m=0}^{\infty}\left(\frac{1+\gamma^{m}}{2}\right)\ketbra{\Phi_m}{\Phi_m}\otimes \ketbra{+}{+} + \sum_{m=0}^{\infty}\left(\frac{1-\gamma^{m}}{2}\right)\ketbra{\Phi_m}{\Phi_m}\otimes \ketbra{-}{-} ~,
\end{equation}
so that the above equation can be rewritten as
\begin{equation}
\label{eq:joint_state}
    \omega_{RBC} = q_+ \omega^{+}_{RB} \otimes \ketbra{+}{+} +q_- \omega^{-}_{RB} \otimes \ketbra{-}{-}~,
\end{equation}
where the conditional states $\omega^{\pm}_{RB}:=\tilde{\omega}^{\pm}_{RB}/q_{\pm}$ are given by
\begin{equation}
\label{eq:joint_state2}
    \tilde{\omega}^{\pm}_{RB}:= \sum_{m=0}^{\infty}\left(\frac{1\pm\gamma^{m}}{2}\right)\ketbra{\Phi_m}{\Phi_m}~, \quad q_{\pm}:=\mbox{Tr}(\tilde{\omega}^{\pm}_{RB})~.
\end{equation}
Equation (\ref{eq:joint_state}) shows that Bob can measure $C$ in this basis without disturbing the unconditional joint state. The measurement outcome $c\in\{+,-\}$ occurs with probability $q_c$ and leaves Alice and Bob sharing the conditional bipartite state $\omega_{RB}^{c}$.
We now construct an achievable two-way assisted quantum communication protocol. Alice prepares independent copies of $\ket{\Psi_N}$ and transmits one mode of each copy through the switch. Bob measures the corresponding control systems and communicates the outcomes to Alice through the available classical channel. For each outcome $c$, the parties can then perform one-way entanglement distillation on the corresponding conditional states $\omega_{RB}^{c}$, using classical communication from Bob to Alice.

Indeed, the hashing inequality \cite{devetak2005distillation}, applied with the roles of Alice and Bob interchanged, guarantees that the one-way distillable entanglement satisfies
\begin{equation}
\label{eq:hashing}
D_{\leftarrow}(\rho_{RB})
\geq
H(R)_{\rho} - H(RB)_{\rho},
\end{equation}
where $D_{\leftarrow}$ denotes the distillable entanglement assisted by one-way classical communication from Bob to Alice. Applying Eq.~(\ref{eq:hashing}) to the conditional states yields
\begin{equation}
D_{\leftarrow}(\omega_{RB}^{c})
\geq
H(R)_{\omega^c}
-
H(RB)_{\omega^c}
\end{equation}
To determine the overall rate, consider $n$ independent uses of the switch. Since Bob communicates the outcomes to Alice, the parties can group the corresponding systems and apply independent reverse-hashing protocols to the $n_c$ copies of $\omega_{RB}^{c}$.
It follows that the achievable entanglement-distillation rate per switch use is at least
\begin{align}
R_{\mathrm{dist}}
\geq
\sum_{c=\pm}q_c
\left[H(R)_{\omega^c}
-
H(RB)_{\omega^c}\right]
\end{align}
The distilled maximally entangled pairs can subsequently be used to teleport quantum states from Alice to Bob. Since teleportation requires only the distilled entanglement and classical communication, without additional uses of the switch, any achievable distillation rate is also an achievable rate for two-way assisted quantum communication. Consequently,
\begin{equation}
\mathcal{Q}_{2,N}(\mathcal S_\lambda)\geq
\sum_{c=\pm}q_c
\left[
H(R)_{\omega^c}
-
H(RB)_{\omega^c}
\right]~,
\end{equation}
where $\mathcal{Q}_{2,N}$ denotes the two-way quantum capacity with mean input photon number at most $N$ per channel use.
Because the control system is classical, we have
\begin{align}
&H(RC)_\omega= H(\{q_c\})+\sum_{c=\pm}q_c H(R)_{\omega^c}~,\\
&H(RBC)_\omega= H(\{q_c\})+\sum_{c=\pm} q_c H(RB)_{\omega^c}~.
\end{align}
Taking the difference between the above entropies, we easily find
\begin{equation}
H(RC)_\omega - H(RBC)_\omega = \sum_{c=\pm} q_c \left[
H(R)_{\omega^c} - H(RB)_{\omega^c}
\right]~,
\end{equation}
from which we derive the following lower bound:
\begin{equation}
\label{eq:two_way_rate}
\mathcal{Q}_{2,N}(\mathcal S_\lambda)
\geq
H(RC)_\omega-H(RBC)_\omega~.
\end{equation}

To justify the application of the hashing inequality in the infinite-dimensional setting while preserving the energy constraint, we follow the approach used in \cite{falco2026energy}. Let $x := N/(N+1)$ and introduce the truncated TMSV state $\ket{\Psi_{N,L}}$ given by
\begin{equation}
\ket{\Psi_{N,L}}
:=
\frac{1}{\sqrt{s_L}}
\sum_{n=0}^{L}\sqrt{t_n}\ket{n}_R\ket{n}_A~,
\quad
s_L:=1-x^{L+1}~.
\end{equation}
The mean input photon number of the state $\tau_{N,L}:= \mbox{Tr}_{R}(\ketbra{\Psi_{N,L}}{\Psi_{N,L}})$ can be computed evaluating the truncated geometric sum. We find
\begin{align}
\label{eq:trunc_input}
N_L:=\mbox{Tr}(\hat{n}_A \tau_{N,L}) = \frac{1}{s_L}\sum_{n=0}^{L} n t_n
= N-\frac{(L+1)x^{L+1}}{1-x^{L+1}}
\leq N~,
\end{align}
which converges to $N$, in the limit $L \to \infty$. Thus, every truncated input satisfies the original energy constraint.

Let us now consider the state $\omega_{RBC}^{(L)}$ given by
\begin{equation}
\omega_{RBC}^{(L)}:= (\mathds{1}_{R} \otimes \mathcal{S}_{\lambda})
(\ketbra{\Psi_{N,L}}{\Psi_{N,L}})~.
\end{equation}
Since this state has finite-dimensional support, the previous reverse-hashing argument applies, giving
\begin{equation}
\label{eq:trunc_lower_bound}
\mathcal{Q}_{2,N}(\mathcal S_\lambda)
\geq
H(RC)_{\omega^{(L)}}-H(RBC)_{\omega^{(L)}}~.
\end{equation}
Adding and subtracting $H(R)_{\omega^{(L)}}- H(RB)_{\omega^{(L)}}$ to the right-hand side of Eq.~(\ref{eq:trunc_lower_bound}) yields
\begin{align}
    \mathcal{Q}_{2,N}(\mathcal S_\lambda)
&\geq
H(RC)_{\omega^{(L)}}-H(RBC)_{\omega^{(L)}} + H(R)_{\omega^{(L)}}- H(RB)_{\omega^{(L)}} -H(R)_{\omega^{(L)}} + H(RB)_{\omega^{(L)}}\\
&= H(R)_{\omega^{(L)}}- H(RB)_{\omega^{(L)}} + H(B|R)_{\omega^{(L)}} + H(C|R)_{\omega^{(L)}} - H(BC|R)_{\omega^{(L)}}\\
\label{eq:low_bound}
&= H(R)_{\omega^{(L)}}- H(RB)_{\omega^{(L)}} + I(B;C|R)_{\omega^{(L)}}~.
\end{align}

\noindent Regarding the first two terms in Eq.~(\ref{eq:low_bound}), in \cite{falco2026energy} it was already shown that
\begin{equation}
    \lim_{L \rightarrow\infty} [H(R)_{\omega^{(L)}}- H(RB)_{\omega^{(L)}}] = H(R)_{\omega}- H(RB)_{\omega} = g(N)-g((1-\lambda^2)N)= \mathcal{Q}_{2,N}(\mathcal{E}_{\lambda^2})~.
\end{equation}

It remains to establish the convergence of the conditional mutual information. This can be done using the following result (see Lemma 2 in \cite{winter2016tight}):
\begin{theorem}
\label{th:winter}
    Let $\rho, \sigma $ be two states on a bipartite Hilbert space $\mathcal{H}_{A} \otimes \mathcal{H}_{B}$. If $T(\rho,\sigma)\leq \epsilon \leq 1$, then
    \begin{equation}
    \label{eq:theo_winter}
        |H(A|B)_{\rho}- H(A|B)_{\sigma}| \leq 2\epsilon \log_{2}(d_{A}) + (1+\epsilon) h_{2}\left(\frac{\epsilon}{1+\epsilon}\right)~,
    \end{equation}
where $d_{A}:= \dim(\mathcal{H}_{A}) < \infty$ and $T(\rho,\sigma):=\frac{1}{2}\norm{\rho-\sigma}_{1}$ is the trace distance between two states.
\end{theorem}

To apply Theorem \ref{th:winter}, let us begin by computing the trace distance $T(\omega_{RBC}^{(L)},\omega_{RBC})$. To this end, note that   $\braket{\Psi_N|\Psi_{N,L}}=\sqrt{s_L}$. Hence, contractivity of the trace distance under quantum channels gives
\begin{equation}
    \frac{1}{2} \norm{\omega_{RBC}^{(L)}-\omega_{RBC}}_1
    \leq \frac{1}{2}\norm{\ketbra{\Psi_{N,L}}{\Psi_{N,L}}-\ketbra{\Psi_{N}}{\Psi_{N}}}_1 =\sqrt{1-s_L}:=\epsilon_{L} \leq 1~.
\end{equation}
Then, Theorem \ref{th:winter} implies
\begin{align}
    & |H(C|RB)_{\omega^{(L)}}- H(C|RB)_{\omega}| \leq 2\epsilon_{L}  + (1+\epsilon_{L}) h_{2}\left(\frac{\epsilon_{L}}{1+\epsilon_{L}}\right)~.
\end{align}
Since the right-hand side vanishes for $L \rightarrow \infty$, we can conclude that, in the same limit, $H(C|RB)_{\omega^{(L)}}$ approaches $ H(C|RB)_{\omega}$. It is immediate to prove an analogous result for $H(C|R)_{\omega^{(L)}}$, so that
\begin{align}
    &\lim_{L \rightarrow \infty}H(C|RB)_{\omega^{(L)}} = H(C|RB)_{\omega}~,\\
    &\lim_{L \rightarrow \infty}H(C|R)_{\omega^{(L)}} = H(C|R)_{\omega}~,
\end{align}
from which we finally find
\begin{equation}
\label{eq:qc_2n}
    \mathcal{Q}_{2,N}(\mathcal S_\lambda) \geq \mathcal{Q}_{2,N}(\mathcal E_{\lambda^2})
    +I(B;C|R)_\omega~.
\end{equation}

Strong subadditivity implies $I(B;C|R)_\omega \geq 0$. To establish strict positivity we proceed by contradiction and assume that $I(B;C|R)_\omega=0$. As a consequence, Proposition 13 in \cite{shirokov2016measures} implies the existence of a CPTP recovery map $\mathcal R_{R\to RC}$ such that $\omega_{RBC} =(\mathcal R\otimes\mathds{1}_B)\omega_{RB}$, with $\omega_{RB}=(\mathds{1}_{R}\otimes \mathcal{E}_{\lambda^2})\ketbra{\Psi_N}{\Psi_N}$.
Let us now compute $\tilde{\rho}^{(k)}:={}_B\bra{k}\omega_{RB}\ket{k}_B$, which corresponds to the (unnormalized) state of $R$, conditioned on Bob measuring $k$ photons. Direct calculation yields
\begin{align}
\tilde{\rho}^{(k)} &= \sum_{n,n'=0}^{\infty} \sqrt{t_n t_{n'}}\ketbra{n}{n'} \bra{k}\mathcal{E}_{\lambda^2}(\ketbra{n}{n'})\ket{k}\\
&= \sum_{n,n'=0}^{\infty} \sqrt{t_n t_{n'}}\ketbra{n}{n'} \bra{k}\left(\sum_{l=0}^{\min(n,n')} \sqrt{\binom{n}{l} \binom{n'}{l}}(1-\lambda^2)^l \lambda^{n+n'-2l} \ket{n-l}\bra{n'-l} \right)\ket{k}\\
&= \sum_{n=k}^{\infty} t_n  \binom{n}{k} (1-\lambda^2)^{n-k} (\lambda^2)^{k}\ketbra{n}{n}~,
\end{align}
where we have used the fact that the action of the pure-loss channel $\mathcal{E}_{\lambda^2}$ on arbitrary Fock states is given by (see, e.g., Lemma 13 in \cite{mele2024quantum})
\begin{equation}
    \mathcal{E}_{\lambda^2}(\ketbra{n}{n'})=\sum_{l=0}^{\min(n,n')}\sqrt{\binom{n}{l} \binom{n'}{l}}(1-\lambda^2)^l \lambda^{n+n'-2l} \ketbra{n-l}{n'-l}~.
\end{equation}
Let us now compute $p_k:=\mbox{Tr}(\tilde{\rho}^{(k)})$, i.e.,
\begin{align}
    p_k&= \sum_{n=k}^{\infty} t_n  \binom{n}{k} (1-\lambda^2)^{n-k} (\lambda^2)^{k}\\
    &=\sum_{l=0}^{\infty} \frac{N^{l+k}}{(1+N)^{l+k+1}}  \binom{l+k}{k} (1-\lambda^2)^{l} (\lambda^2)^{k}\\
    &=\frac{(N\lambda^2)^k}{(1+N)^{k+1}} \sum_{l=0}^{\infty} \left(\frac{N(1-\lambda^2)}{1+N}\right)^l  \binom{l+k}{k}\\
    \label{eq:p_k}
    &=\frac{(N\lambda^2)^k}{(1+N)^{k+1}} \sum_{l=0}^{\infty}  \binom{l+k}{k} z^l~,
\end{align}
where we have set $z:=\frac{N(1-\lambda^2)}{1+N}$. 
The last term in the above equation can be simplified by starting from the geometric series
\begin{equation}
    \sum_{l=0}^{\infty}  z^l = \frac{1}{1-z}~,
\end{equation}
which converges for $|z|<1$. Differentiating both sides $k$ times and dividing by $k!$ yields
\begin{equation}
    \sum_{l=0}^{\infty}  \binom{l+k}{k} z^l = \frac{1}{k!} \frac{d^k}{dz^k}\left(\frac{1}{1-z}\right) = \frac{1}{(1-z)^{k+1}}
\end{equation}
Hence, Eq.~(\ref{eq:p_k}) becomes
\begin{equation}
    p_k= \frac{(N\lambda^2)^k}{(1+N)^{k+1}} \frac{1}{(1-z)^{k+1}}=\frac{(N\lambda^2)^k}{(1+N \lambda^2)^{k+1}}~,
\end{equation}
from which we can define the normalized state $\rho^{(k)}:=\tilde{\rho}^{(k)}/p_k$, given by
\begin{equation}
    \rho^{(k)} = \frac{1}{p_k}\sum_{n=k}^{\infty} t_n  \binom{n}{k} (1-\lambda^2)^{n-k} (\lambda^2)^{k}\ketbra{n}{n} = \sum_{n=k}^{\infty} \rho^{(k)}_{n}\ketbra{n}{n}~,
\end{equation}
where we have set $\rho^{(k)}_{n}:=\frac{1}{p_k} t_n \binom{n}{k} (1-\lambda^2)^{n-k} (\lambda^2)^{k}$.
The action of the recovery map on $\rho^{(k)}$ yields
\begin{equation}
    \mathcal{R}(\rho^{(k)})=\frac{1}{p_k}\mathcal{R}({}_B\bra{k}\omega_{RB}\ket{k}_B) = \frac{1}{p_k}{}_B\bra{k}\omega_{RBC}\ket{k}_B~,
\end{equation}
from which we find
\begin{equation}
    \mathcal{R}(\rho^{(k)})=\sum_{n=k}^{\infty}\rho^{(k)}_n \ketbra{n}{n} \otimes \sigma_{n-k}~.
\end{equation}
Let us now compute the fidelity between the states $\rho^{(0)}$ and $\rho^{(1)}$, which reads
\begin{equation}
    F(\rho^{(0)},\rho^{(1)})=\left(\mbox{Tr}\sqrt{\sqrt{\rho^{(0)}} \rho^{(1)} \sqrt{\rho^{(0)}}} \right)^2 = \left( \sum_{n=1}^{\infty} \sqrt{\rho_n^{(0)} \rho_n^{(1)}} \right)^2~.
\end{equation}
An analogous calculation yields
\begin{equation}
     F\left(\mathcal{R}(\rho^{(0)}), \mathcal{R}(\rho^{(1)}) \right) = \left( \sum_{n=1}^{\infty} \sqrt{\rho_n^{(0)} \rho_n^{(1)}} \sqrt{F(\sigma_{n}, \sigma_{n-1})} \right)^2~,
\end{equation}
where we have used the formula for the fidelity between two classical-quantum states (see, e.g., Eq.~(9.146) in \cite{wilde2017quantum}). Taking the term $n=1$, we easily find
\begin{equation}
    F(\sigma_{0}, \sigma_{1}) = \braket{+|\sigma_1|+} =\frac{1+\gamma}{2}~.
\end{equation}
Restricting to $0 < \lambda<1$, it follows that $F(\sigma_{0}, \sigma_{1}) <1$, from which we can conclude that
\begin{equation}
     F\left(\mathcal{R}(\rho^{(0)}), \mathcal{R}(\rho^{(1)}) \right)  < F(\rho^{(0)},\rho^{(1)})~.
\end{equation}
However, monotonicity of fidelity under CPTP maps requires that $F\left(\mathcal{R}(\rho^{(0)}), \mathcal{R}(\rho^{(1)}) \right)  \geq  F(\rho^{(0)},\rho^{(1)})$, thus contradicting our initial assumption $I(B;C|R)_\omega=0$. Hence, $I(B;C|R)_\omega>0$ for every $N>0$ and $0<\lambda<1$. Then, Eq.~(\ref{eq:qc_2n}) implies
\begin{equation}
    \mathcal{Q}_{2,N}(\mathcal{S}_\lambda) > \mathcal{Q}_{2,N}(\mathcal{E}_{\lambda^2})~,
\end{equation}
thus concluding our proof.

\bibliographystyle{apsrev4-2}
\bibliography{ref.bib}
\end{document}